\documentclass[prd,12pt,tightenlines,nofootinbib]{revtex4}
\usepackage{amsfonts,amssymb,amsthm}

\usepackage{amsmath,amssymb}

\newcommand{\be}{\begin{equation}}
\newcommand{\ee}{\end{equation}}
\newcommand{\beq}{\begin{eqnarray}}
\newcommand{\eeq}{\end{eqnarray}}
\newcommand{\bea}{\begin{eqnarray}}
\newcommand{\eea}{\end{eqnarray}}

\newcommand{\bra}{\langle}
\newcommand{\ket}{\rangle}

\newcommand{\rd}{\mathrm{d}}
\newcommand{\tr}{{\rm tr}}

\newcommand{\bz}{\bar{z}}
\newcommand{\tu}{\tilde{u}}

\newcommand{\tmu}{\tilde{\mu}}

\newcommand\hpi{{\hat{\Pi}}}
\newcommand\hK{{\hat{K}}}
\newcommand\bga{\gamma}

\begin{document}
\title{Reconstructing AdS/CFT}



\author{Laurent Freidel}
\affiliation{Perimeter Institute for Theoretical Physics,
31 Caroline Street North, Waterloo, N2L 2Y5, Canada.}

\date{\small \today}

\begin{abstract}
In this note we  clarify the dictionary  between pure quantum gravity on the bulk in the presence of a cosmological constant and 
a CFT on the boundary.
We show for instance that there is a general correspondence between 
quantum gravity ``radial states'' and a pair of CFT's.
Restricting to  one CFT is argued to correspond to states possessing an asymptotic infinity.
This point of view allows us to address the problem of reconstructing the bulk from the boundary.
And in the second part of this paper we present an explicit formula which gives, from the partition function of any 2 dimensional conformal field theory,
a wave functional solution to the 3-dimensional Wheeler-DeWitt equation. This establishes at the quantum level a precise dictionary between 2d CFT and pure gravity.
\end{abstract}

\maketitle
\section{Introduction}
 AdS-CFT is a deep and fascinating correspondence between  a bulk theory of gravity and a boundary Conformal Field Theory
 which was first conjectured in the context of string theory \cite{maldacena1}.
The purpose of this note is to study from the point of view of quantum gravity  the relationship between a theory of pure quantum gravity in the bulk  and a boundary CFT.
The main question we want to address here is the question : What is AdS/CFT from the point of view of background independent quantum gravity?

Trying to answer this  leads to many related questionings.
An obvious one is: Assuming that we have a theory of quantum gravity in the bulk, {how can we prove the AdS/CFT correspondence?}
For instance, in 3 dimension a large body of work has led us to 
the conviction that 3d pure quantum gravity is a theory which 
can be consistently defined\footnote{ By this we mean that we can give a definition of the path integral, define certain transition amplitude, some set of observables and solve the Wheeler-DeWitt equation, at least in the context of compact manifolds and also in the presence of particles \cite{PR}. There have been a wealth of techniques developed from Chern-Simons quantization, spin foam approach, t'Hooft approach etc... which  have somehow different range of applicability 
but  agree when they can be applied to the same problem (see \cite{Carlip} and ref therein).}.
Can we then prove or disprove the recent proposal made by Witten for a CFT dual to pure 3d gravity \cite{Witten3d}? 
Clearly, answering this question amounts to ask what is the precise dictionary between AdS Quantum gravity and boundary CFT?

It is sometimes claimed that the CFT provides a {\it definition} of what 
a background independent quantum gravity theory  in the bulk should be or that solving quantum gravity amounts to find the CFT it is dual to
 \cite{Witten3d}. Is it true?  If so is there a way to reconstruct the bulk from the boundary?
How? What should be the equation driving the reconstruction?

Concerning the correspondence one may ask: Is the correspondence one to one between one theory of QG to one CFT? or is it one to many? If so, how can we talk about {\it the} CFT associated to quantum gravity? And how can we think of proving the correspondence?

Another concern sometimes raised is that gravity is background independent, is it in conflict or not with restricting  to asymptotically AdS space time? Can we formulate such a restriction starting from a background independent approach of quantum gravity?

Finally, one may wonder:
Is AdS/CFT a property of string theory in certain backgrounds only or is it a genuine and intrinsic property of any theory of quantum gravity
whatever is its realization?

There is a vast literature on the subject of AdS/CFT, however we feel that only a relative small portion of it focuses on  the questions asked here :
 First there is the key original reference of Witten \cite{Witten1} where 
 the meaning of the correspondence is proposed; the body of work 
 by Skenderis et al. \cite{Skenderis1, Skenderis2, HJ2} developing the semi-classical understanding of the correspondence and the holographic renormalisation group; the central claim  of Verlinde's and deBoer \cite{deBoer} that there is a relationship between the Hamilton-Jacobi equation and the holographic renormalisation group,
 and a  Schr\"odinger picture sketched  by Maldacena \cite{Maldacena} in the de Sitter context.
 
Despite these key references, it is hard to  find one reference addressing all the questions asked here in a systematic fashion. 
One of the purpose of this note is to  try to fill this gap and present a unified interpretation of the AdS/CFT correspondence from the Quantum gravity point of view. We hope  that even specialists will learn something new about AdS/CFT from our presentation.
 
 In the second part of this paper and using the understanding of AdS/CFT developped in the first part we give an explicit reconstuction formula of the bulk amplitude given a CFT in the 2+1 dimensional case.

In section \ref{sec1} of this note we present some well known fact about AdS/CFT and the original formulation of the AdS/CFT correspondence.
We then identify three problems with this original formulation and study 
independently the equations satisfied by the bulk gravity  and the boundary CFT.
  In section \ref{sec2} we show how these puzzles 
  can be resolved by looking at the asymptotics of the gravity amplitude. And we reach several key and sometimes unusual conclusions about the AdS/CFT correspondence in section \ref{seckey}.
  In section \ref{bulkrec} we present the general features, in any dimension, of the reconstruction of the bulk from the boundary
  CFT.
  Moreover we give, in dimension 2+1, an explicit  reconstruction formula (\ref{main}) for the quantum gravity amplitude given {\it any} boundary CFT.
 In the last sections \ref{1order}, \ref{proof} we give the proof of this  reconstruction formula.
  

\section{On the AdS-CFT Correspondence}
\label{sec1}
We want in this section to review some known fact about this correspondence and clarify some puzzles concerning  the precise dictionary expected between AdS and CFT .

At the semi-classical level this correspondence can be stated quite precisely \cite{Witten1,Polgub} by looking at space-times which are asymptotically AdS and  by relating the value of the classical on-shell action of bulk fields with the expectation value of corresponding operators in the CFT.
The main focus of this paper is the case of Lorentzian AdS manifolds, but some of the results in the following section can be also stated for 
Lorentzian dS manifolds, so we will consider both cases simultaneously.

Suppose that we have a background Lorentzian spacetime $M,g$ of dimension $d+1$ solution of 
Einstein equations with a cosmological constant $\Lambda$ 
$$R_{\mu\nu}(g) = - \epsilon \frac{d}{\ell^{2}}g_{\mu\nu},\quad\,\,\Lambda=-\epsilon \frac{d(d-1)}{2\ell^{2}}$$ with $\epsilon=-1$ for dS and $\epsilon= +1$ for AdS. 
This manifold is conformally compact if there exists a defining function $\rho$ such that   $\rho^{-1}(0)= \partial M$ and 
$d\rho\neq 0$ on $\partial M $ and such that the conformally equivalent metric 
\be
\ell^{2} \bar{g} = \rho^{2} g
\ee
extends smoothly to  the boundary of $M$.
The Einstein equations imply that \be\label{defining}
 \bar{g}^{\mu\nu}\partial_{\mu}\rho\partial_{\nu}\rho = \epsilon \ee on $\partial M$. The conformal infinity is therefore Lorentzian in AdS and Riemannian in dS. 
It is always possible to chose a defining function $\rho$ such that the relation (\ref{defining}) is valid not only on $\partial M$ but also in a neighborhood of $\partial M$. 
For instance, when $M$ is conformally compact  we can take as a defining function $\ell \rho(x)=d_{\bar{g}}(\partial M,x)$, the distance of $x$ from $\partial M$ with respect to $\bar{g}$.
In this coordinates  the metric can be written
\be
ds^{2} = \frac{\ell^{2}}{\rho^{2}} (\epsilon d \rho^{2} + \gamma_{\rho}),\quad ds^{2} =   (\epsilon d r^{2} + \ell^{2} e^{\frac{2r}{\ell}}\gamma_{r}),\quad \rho=\exp\left(-\frac{r}{\ell}\right)
\ee
where $\rho^{-2}\gamma_{\rho}$ is the metric induced on the 
surfaces $\Sigma_{\rho}$ with $\rho=cste$ 
and $\rho=0$ defines the conformal infinity of $M$.
The surfaces $\Sigma_{\rho}$ are spacelike in the dS case and timelike in AdS case. In both cases $\epsilon= n^{2}$ is the 
square norm of the normal vector to $\Sigma_{\rho}$.
In the AdS case $r$ is the radial geodesic distance for the physical metric $g$.

The normal vector and extrinsic curvature of the surfaces $\Sigma_{\rho}$ are given by 
\be
n= \partial_{r}=-\frac{\rho}{\ell}\partial_{\rho},\quad K_{\mu}^{\nu} = h^{\nu \alpha}\nabla_{\alpha}n_{\mu} =\frac12 h^{\nu\alpha}{\cal L}_{n}g_{\alpha \mu}
\ee
where $h_{\mu\nu} =g_{\mu\nu} -\epsilon n_{\mu}n_{\nu}$.
This extrinsic curvature can be easily computed in our coordinates and this gives
\be
\label{excurv}
\ell K_{i}^{j} = \left(\delta_{i}^{j} - \rho  (\gamma^{-1}_{\rho}\partial_{\rho}\gamma_{\rho})_{i}^{j}\right) = \delta_{i}^{j} + O(\rho^{2}).
\ee
One can easily see that $\partial_{\rho}\gamma|_{\rho=0}=0$ thus the extrinsic curvature tensor is proportional to the identity operator up to order $\rho^{2}$ at infinity, this fact plays a key role in the AdS-CFT correspondence even at the quantum level.
In order to see why lets consider wave function depending on the 
metric $\bga_{ij}$ on $\Sigma_{\rho}$ and lets consider a radial evolution\footnote{Of course this evolution is timelike in the dS case}
of this wave function towards the conformal infinity
\be
\partial_{r} \Psi(\gamma_{\rho}) =\int_{\Sigma_{\rho}} \partial_{r} \bga_{ij} \frac{\delta\Psi}{\delta \bga_{ij}} \sim_{\rho=0} \frac2{\ell}\int_{\Sigma_{0}}  \bga_{ij}  \frac{\delta\Psi}{\delta \bga_{ij}}.
\ee
The last operation becomes just the operation of conformal rescaling.
Thus thanks to the presence of the cosmological constant a radial evolution in the bulk is equal near infinity  to a conformal transformation at the  boundary.
Clearly, infinity is left invariant by the radial evolution and therefore one expect conformal invariance of the physics described by $\Psi$ at asymptotic infinity. As we will see more precisely, this is the essence of the correspondence between bulk gravity and boundary CFT.

The AdS/CFT correspondence as originally stated by Witten \cite{Witten1,Polgub} is  a relation between the partition function of quantum gravity with fixed boundary data in the bulk and the generating functional of connected correlation functions of the CFT on the boundary.
Namely Lets $\Phi_{i}= \Phi,g_{\mu \nu},...$ denote fields that propagate in the bulk  and lets denote the asymptotic boundary value of this field on $\Sigma_{\rho}$ by $\phi_{i}$.
On the gravity side, one can define the  amplitude
\be\label{S-matrix}
\Psi_{{\Sigma_{\rho}}}(\phi_{i}) =\int_{\Phi_{i}|_{\Sigma_{\rho}}= \phi_{i}}\!\!\!\!\!\!\!\mathrm{D} \Phi_{i} \,\,e^{i S_{B,M}(\Phi_{i})}  
\ee
Where $S_{B,M}$ is the bulk action and the bulk partition function is evaluated with Dirichlet boundary condition on the fields.
It is important to note that if we where in asymptotically flat space,
asymptotic infinity would be null  ($\partial M = {\cal I}^{+}\cup {\cal I}^{-}$ in flat space) but we can still define the same type of Dirichlet amplitudes if we go a bit away from infinty and the 
 take a spacelike slice.
What has been established a long time ago \cite{Faddeev} but which is  not often stated explicitly is that this 
Dirichlet amplitude is, in this case, nothing but the S-matrix functional.
The usual S-matrix elements can be obtained from this functional 
by taking derivative of the S-matrix.
It is then tantalizing to call the amplitude (\ref{S-matrix}) the 
``(Ad)S-matrix functional''.
This object is naturally related to the quantum effective action
$\Gamma(\Phi_{i})$ which depends on the bulk fields and is obtained by performing the path integral in the presence of background bulk fields $\Phi_{i}$.
Given the quantum effective action $\Gamma(\Phi_{i})$ we evaluate it on-shell and  compute its Hamilton-Jacobi functional 
\be
S(\phi_{i}) \equiv \left.\Gamma(\Phi_{i})\right|_{\frac{\delta\Gamma}{\delta \Phi_{i}}=0,\,\,\Phi_{i}|_{\partial M}= \phi_{i}}.
\ee 
This quantum Hamilton-Jacobi functional is then (in flat space)  the generating functional for  connected S-matrix elements: $\Psi(\phi_{i})= e^{iS(\phi_{i})}$, the bulk amplitude is thus an {\it on-shell} amplitude.

On the other hand, on the CFT side one associates to each boundary field $\phi_{i}$ 
a primary operator $\hat{O}_{i}$ of the CFT.
Different fields $\phi_{i}$ are characterized not only by their tensorial structure but also by their properties under conformal transformation of the boundary metric. Namely when $\gamma \to \rho^{-2}\gamma$ the fields transform as 
$\phi_{i}\to \rho^{d-\Delta_{i}}\phi_{i}$. 
$\Delta_{i}$ is then the conformal dimension of the operator $O_{i}$.
Given the CFT one can define the generating functional of connected correlation functions
\be
Z_{CFT}(\phi_{i}) = \bra e^{\int_{\Sigma} \phi_{i}\hat{O}_{i}}\ket
\ee
Note that this is an {\it off-shell} amplitude.
The AdS/CFT correspondence is the statement that there is an equality
\be\label{equal}
\Psi_{{\Sigma_{0}}}(\phi_{i})= Z_{CFT}(\phi_{i}).
\ee

\subsection{Three puzzles}

This original formulation is however not precise enough and leads to several puzzles.
Formulating and resolving these puzzles is a key part of establishing a proper understanding of the deep nature of AdS/CFT correspondence 
and  will allow us to propose a  bulk reconstruction formula.
We can clearly identify  three main problems with the original formulation:

{\bf 1-}The first puzzle is purely technical in nature and has been identified since the beginning \cite{Witten1}, it comes from the fact that 
 the evaluation of the (Ad)S-matrix is at conformal infinity (it is computed by taking a limit $\rho\to 0$ towards asymptotic infinity) 
 there are, even at the classical level, infinities arising in its evaluation that needs to be taken care of and substracted.

{\bf 2-}The second puzzle is more conceptual but as important as the previous one. It comes from the fact that the formulation given here is in term of a background metric
since it usually explicitly refers to the spacetime slicing given by $\rho$.
We will be interested in quantum gravity, and in quantum gravity the metric is a dynamical object  and cannot be fixed beforehand. 
Even more, in quantum gravity the quantum spacetime is represented by the knowledge of the wave function $\Psi$.
 The formulation of the correspondence should therefore 
be independent of a choice of a background metric. But then where is the asymptotic boundary?

{\bf 3-}Finally, and this is  at first sight the most serious puzzle, the two objects in (\ref{equal}) do {\it not} satisfy the same equations!
One is a solution of Wheeler-deWitt equation which is a second order differential equation and the other a solution of a conformal Ward identity which is a first order equation.
How  could one have equivalence between a second and first order differential system?

Hopefully this three puzzles are related and can be resolved all together using 
the idea of  the so called ``holographic renormalisation group'' which is thus a key and central ingredient 
for the precise  formulation of the AdS/CFT correspondence.
We review here some of the results of the holographic renormalisation group \cite{Skenderis1, Skenderis2} but we hope to give a new and fresh perspective on some of the results and on the resolutions of these puzzles that 
 will allows us to go further.

Lets us start to analyse first what type of equations $\Psi$ and $Z_{CFT}$ are supposed to satisfy. We will look from now on to the case of pure gravity.

\subsection{Gravity equations}\label{grav}
Let us first analyse the gravity sector. In this case we are interested in  the following functional of a metric $\bga$ on a d dimensional space $\Sigma$.
In the case this manifold is the topological boundary of a d+1 dimensional manifold $M$, $\partial M =\Sigma$, we consider
\be\label{SG-matrix}
\Psi_{{\Sigma}}(\bga) =\int_{g|_{\partial M}= \bga}\!\!\!\!\!\!\!\mathrm{D} g \,\,e^{i S_{M}(g)}  
\ee
This is indeed a formal expression and the  problem of quantum gravity 
is  to make sense of it. 
One can hope that there is a precise definition of this object in string theory or non perturbative gravity which is consistent.
In our case, we will later  work in $2+1$ gravity where a non perturbative 
definition of Lorentzian quantum gravity and proposal for this partition function exists. 
In higher dimension 
we can also think about this integral to be defined at one loop  in which case it is perfectly meaningful,
keeping in mind that we expect any theory of quantum gravity to be in agreement with the one loop results.

In order to write down  the Einstein action associated with a  $d+1$ dimensional manifold $M$
having a boundary $\Sigma$,  lets introduce some notations. 
We denoted by $g_{\mu\nu}$ the metric on $M$, $n^{\mu}$ the unit vector normal to the boundary $\Sigma$. We have that $n^{2} = -1$ if  $\Sigma$ is spacelike (dS case)
and   $n^{2}=+1$, if  $\Sigma$ is timelike (AdS case). In both case we denote $\epsilon= n^{2}$.
We also denote
$\gamma_{\mu \nu}= g_{\mu \nu} - \epsilon n_{\mu}n_{\nu}$ the boundary metric  on $\Sigma$ and 
\be
K_{\mu \nu} = \bga_{\mu}^{ \rho} \nabla_{\rho} n_{\nu}=\frac12 {\cal L}_{n}\bga_{\mu \nu}
\ee
is the extrinsic curvature tensor.
The action is 
\be
S_{M}(g)=- \left(\frac1{2\kappa} \int_{M} {\mathrm{d}}^{d+1}x \sqrt{g} \left(R(g) -2\Lambda\right) +  \frac{\epsilon}{\kappa}\int_{\Sigma}  {\mathrm{d}}^{d}x \sqrt{\bga}K  \right)
\ee 
where\footnote{Our conventions are that the metric signature is $(-+++)$ and the curvature tensor is given by
$R_{\mu\nu\alpha}^{\beta} = - \partial_{\mu} \Gamma_{\nu \alpha}^{\beta} - \Gamma_{\mu \sigma}^{\beta}\Gamma_{\nu \alpha}^{\sigma}
+ (\mu \leftrightarrow \nu)$. This convention ensure that the Euclidean sphere as a positive scalar curvature, and that AdS has a negative scalar curvature.}
 $$\Lambda = -\epsilon \frac{d(d-1) }{2 l^{2}}, \quad \kappa \equiv 8\pi G, \quad K=K_{\mu \nu} \bga^{\mu \nu}$$
and $l$ is the cosmological scale.
The boundary term\footnote{ If the boundary conditions is such that the boundary metric 
vary discontinuously or change signature we need to include additional boundary terms \cite{HawkingH, Hayward} to the action 
of the form $$+ \frac1{2\kappa} \int_{J} \rd^{d-1}\sqrt{\bga} \Theta$$ where $\Theta $ is the angle or boost parameter along the $d-1$ joints
at which the normal change signature or vary discontinuously. }
 is necessary in order to have a well defined 
variational principle \cite{HawkingG}.
Indeed for an on-shell\footnote{ We have chosen the overall sign of $S_{M}$ so that when we couple to matter field 
the total action reads $S_{M} + S_{m}$ for instance for scalar field we take 
$$S_{m}(\phi) =\frac12  \int\sqrt{g} \partial_{\mu}\phi\partial_{\nu}\phi + m^{2}\phi^{2}.$$
The energy momentum tensor being such that 
$\delta S_{m}= \int_{M} \frac{\sqrt{g}}{2} T_{\mu\nu} \delta g^{\mu \nu}$. } variation we have 
\be
\delta S =\frac1{2\kappa} \int_{ \Sigma} \sqrt{\gamma}\, \Pi^{ab } \delta \gamma_{ab },\quad 
\Pi^{ab} = {\epsilon} \left(K^{ab}- \bga^{ab} K \right)\label{pi}
\ee
One can also easily show that under a bulk diffeomorphism $\delta_{\xi}g_{\mu\nu} = {\cal{L}}_{\xi}g_{\mu\nu}$ the action transforms as 
\beq
\delta_{\xi}S_{M}(g) &=&-\frac1{2\kappa} \int_{\partial M } {\mathrm{d}}^{d}x \xi_{n} \left(\sqrt{\bga}
\left(R(g) -2\Lambda\right)  + 2{\cal L}_{n}(\sqrt{\bga}K)  \right) \\
&=&-\frac1{2\kappa} \int_{\partial M } {\mathrm{d}}^{d}x \sqrt{\bga} \xi_{n} \left(
R(\bga) -2\Lambda +\epsilon(K^{2} - K_{a}^{b}K_{b}^{a}) \right)\label{xir}
\eeq
where $\xi_{n}= \xi^{\mu}n_{\mu}$ and in the last equality 
we have used the Gauss-Codazzi equations\footnote{ We mean the identity
\be
{R}(g) + 2\epsilon {\nabla}_{\mu}(n^{\mu} K)
= R(\bga) +\epsilon(K^{2} - K_{a}^{b}K_{b}^{a}) +2\epsilon{\nabla}_{a} a^{a}. 
\ee
where 
$a_{a}={\nabla}_{n} n_{a}$ is the acceleration
and $g_{\mu \nu}= \bga_{\mu \nu} - \epsilon n_{\mu}n_{\nu}$.}
to write the bulk Lagrangian in terms of the boundary metric and extrinsic curvature tensor.
Using the invariance under diffeomorphism of the measure of integration we can
rexpress the change of the action under diffeomorphism by a boundary 
transformation $\delta_{\xi}\bga_{ab} = 2\nabla_{(a}\xi_{b)} + \xi_{n} 2 K_{ab}$.
Using this and (\ref{pi},\ref{xir}) one gets 
the Ward identities expressing the invariance of $\Psi(\bga)$ under bulk
diffeomorphisms.
Lets denote 
$$\hpi^{ab}_{x}\equiv \frac{2}{\sqrt{\gamma}}\frac{\delta\,\,\,\,\,\,\,}{\delta \bga_{ab}(x)},\quad  \hpi^{a}_{b} \equiv \gamma^{ac}\hpi_{cb}, \quad \hpi \equiv \hpi^{c}_{c}$$ 
Note that thanks to (\ref{pi}) the action of the functional derivative on 
$\Psi$ amounts to compute expectation value of ${\cal L}_{n}\gamma$ 
that is 
\be\label{piK}
\hpi^{ab}_{x} \Psi_{\Sigma}(\bga) = \frac{i \epsilon}{\kappa} \bra K^{ab}(x)- \bga^{ab} K(x)\ket_{\gamma} 
\ee
where the expectation value is taken with respect to the gravity measure (\ref{SG-matrix}).
The Ward identity expressing the invariance of $\Psi(\bga)$ under bulk diffeomorphism can be written  ${\cal H}\Psi_{\Sigma} = {\cal H}_{a} \Psi_{\Sigma} =0$ where 
\beq
{\cal H}_{b}&=& \nabla_{a} \hpi^{a}_{b} \\
{\cal H} &=&   -{\epsilon \kappa^{2}} 
:
\left(
\hpi_{a}^{b}\hpi^{a}_{b} -\frac{\hpi^{2}}{d-1}
\right)
: + 
 {R(\bga)}  +\epsilon \frac{d(d-1)}{ l^{2}}
 \label{WdW}
\eeq
This are similar to the usual Hamiltonian constraint equations where ${\cal H}_{b}$ is the generator of infinitesimal diffeomorphism. 
Usually, that is in the hamiltonian picture, these equation are constraint equations which are written only for the case when 
$\gamma$ is a spacelike boundary metric in a Lorentzian manifold.
But as we have just seen seen these equations are Ward identities expressing the invariance of the gravity partition function under  bulk diffeomorphisms, they can be derived even in a generalized context where the boundary is Lorentzian.
In the deSitter case since the boundary surfaces are taken to be spacelike these are exactly the usual 
Hamiltonian equations.
However  in the AdS case (\ref{WdW}) is no longer constraint equations but an evolution equation, 
we will  call  it (\ref{WdW}) the ``radial'' Wheeler-deWitt equation.
Note also that often, what is consider in the literature is the case of Euclidean hyperbolic gravity where $iS_{M}$ is replaced by $-S_{M}$ and $\bga$ is Euclidian.
 We get in this case a similar equation which can be obtained from the AdS radial equation by the ``Wick rotation''  $\hpi^{ab} \to i \hpi^{ab}$, this amounts to change the sign of the kinetic term in  
(\ref{WdW}).

Note however that since in the AdS case we are dealing with the radial Wheeler-deWitt equation, 
this means we do {\it not} expect an a priori relation between Euclidian and Lorentzian gravity in the AdS case.
 This is in stark contrast with the usual case where we can expect some correspondence between Euclidian and Lorentzian solutions of Wheeler-deWitt equation.
Let us expand a bit on this point; and lets  denote the kinetic and potential term  by $T\equiv\hpi_{a}^{b}\hpi^{a}_{b} -\frac{\hpi^{2}}{d-1}$ and $V= R(\bga) -2\Lambda$.
We also denote by $H_{L}$ the usual hamiltonian constraint operator associated with a slicing by constant time surface in a Lorentzian manifold
and by $H_{E}$ the constraint associated to an arbitrary slicing in an Euclidean manifold, these two operators are related since  
$$H_{L} =  \kappa^{2} T + V,\quad \quad  H_{E} = -\kappa^{2} T + V.$$ 
In both cases the state $\Psi_{L,E}(\bga)$ annihilated by $H_{L,E}$ is a functional of a $d$ dimensional Riemannian metric.
Now suppose that $\psi_{E}(\bga)=  \exp(-\frac{1}{\kappa} \Gamma(\bga,{\kappa}))$,  $ \Gamma(\bga,\kappa)\equiv\sum_{n=0}^{\infty}\kappa^{n} \Gamma_{n}(\bga)$,
with $\Gamma_{n}$ real,  is a solution of the Euclidean  hamiltonian
constraints then  $\psi_{L}(\bga)= \exp(\frac{i}{\kappa} \Gamma(\bga,i\kappa))$ is a solution of the Lorentzian constraints.
So there is a clear mapping between the lorentzian and euclidean sector in this case\footnote{In other words $\Psi_{L}(\bga) \equiv \int_{g|_{\Sigma}=\bga}\!\! Dg e^{iS^{(L)}(g)}$ and $\Psi_{E}(\bga) \equiv \int_{g|_{\Sigma}=\bga}\!\!  Dg e^{- S^{(E)}(g)}$
formally satisfy the same WdW equation. 
Here $S^{(L)},(resp. S^{(E)})$ denotes the Lorentzian (resp. Euclidian) action and $\bga$ is a Riemannian
metric.}
This is the reason behind Hartle-Hawking proposal \cite{HawkingHartle}. 

Now in the case of AdS the constraint satisfied by the AdS-matrix functional is of the form 
$$H_{L}^{(radial)} =  - \kappa^{2} T + V,$$
moreover the boundary metric is lorentzian instead of Riemannian,
since it correspond to a radial slicing.
There is no longer, in this case, any simple correspondence between the Lorentzian  and Euclidean solution.
The only correspondence one can naively think of is to analytically continue  $\Psi_{L}(\bga)$ into a functional of non degenerate complex metrics 
and evaluate it on a Riemannian section. Since  $H_{L}^{(radial)}$ and $H_{E}$ are the same this will give a solution of $H_{E}$. However this solution 
will be of the form $\exp(\frac{i}{\kappa} \Gamma(\bga_{E}))$ and not of the form   $\exp(-\frac{1}{\kappa} \Gamma(\bga_{E}))$.
So this will give a solution of Riemannian quantum gravity but  not  of  Euclidian gravity.
One can otherwise  try to perform a Wick rotation $t\to it$ of the action $\Gamma(\bga)$.
Such a rotation can be performed meaningfully if  $\bga$ possesses  at least one timelike killing vector field.
This might give a prescription for $\Psi_{E}(\bga)$ in terms of $\Psi_{L}(\bga)$ for stationary $\bga$.
Since there exists no preferred  timelike Killing vector field for a general $\bga$ it is not really clear how to extend 
this prescription meaningfully to general metrics.
Any such extension would amount to pick a particular gauge and an associated preferred time coordinate.
Even if we do so it is even less clear wether such a extension, if it exists,  maps solutions of $H_{L}$ to solutions of $H_{E}$
(We know that such extension do not generally maps classical euclidean solutions to Lorentzian classical solutions \footnote{Except in three dimensions where it is possible 
to find such extension since Lorentzian and Euclidian classical spacetime  are respectively quotient of $AdS_{3}$ (resp. $H_{3}$)
by discrete subgroup $G$ of $SL(2,\mathbb{R}) \times SL(2,\mathbb{R})$ (resp. $SL(2,\mathbb{C})$). 
Fixing a set of free generators of $G\subset SL(2,\mathbb{R}) \times SL(2,\mathbb{R})$ we can analytically continue these to a set of generator 
of $SL(2,\mathbb{C})$ \cite{Kirill1}.
Such an extension is uniquely defined only for static spacetime but in general there is an infinite number of inequivalent extensions.
Moreover any such extension generally maps many Lorentzian solutions to the same Euclidean solution.
} ).
This lead us to the conclusion that  one cannot  expect beforehand a deep relationship between Euclidean and Lorentzian 
quantum gravity in the AdS/CFT correspondence  as it is often (if not always) assumed.
One may be able to establish such Euclidian/Lorentzian correspondence in some limited regime (perturbation around static boundary space, for instance) 
but so far this is  still an interesting open question.

The dots $:\,:$ in equation (\ref{WdW}) denotes normal ordering terms 
necessary to define the kinetic term of the hamiltonian.
From the derivation of this equation, one sees that the action of the kinetic term on $\Psi(g)$ leads to evaluation of the two point function at coincident point and the associate divergence needs to be subtracted following a prescription initially designed by Symanzik \cite{Symanzik}. 
Namely if we denote by  ${{\cal{G}}^{D}}_{\alpha \beta\mu\nu}(x,y)$ the two point function $ \bra g_{\alpha \beta}(x) g_{\mu\nu}(y)\ket$ 
calculated with Dirichlet boundary\footnote{More precisely we should fix Dirichlet conditions $g_{ab}=\bga_{ab}$  on the metric components tangential to the boundary.
And we should use Neuman conditions induced by the gauge fixing on the other components $g_{0a}$ which depend on the normal direction to the boundary \cite{Barv1}, (see appendix).}
condition and $G^{abcd}= 1/2(\bga^{ac}\bga^{bd} +\bga^{bc}\bga^{ad}) - \bga^{ab}\bga^{cd}$ the boundary supermetric.
One consider the boundary to boundary propagator 
$$
{\cal{K}}^{ab}_{cd}(x,y) = G^{ab mn}
\left.\overrightarrow{\cal{L}}_{n}{\cal{G}}^{D}_{mn cd}(x,y)\overleftarrow{\cal{L}}_{n}\right|_{x,y\in\partial M}
$$
where ${\cal{L}}_{n}$ denotes the Lie derivative with respect to   the normal to the boundary.
Now the boundary to boundary propagator can be split into a part which is singular in the coincident limit and a regular part.
The singular part  ${\cal{K}_{S}}$ of this propagator needs to be subtracted from 
the kinetic term of the hamiltonian in order to get a well defined hamiltonian this is what the normal order stands for. 
\be \label{substraction}
:\left(
\hpi_{a}^{b}\hpi^{a}_{b} -\frac{\hpi^{2}}{d-1}
\right)
: \equiv G_{ab cd}\hpi^{ab}_{x}\hpi^{cd}_{x}  -{\cal{K}_{S}}^{ab}_{ab}(x,x) 
\ee
This Schr\"odinger renormalisation  procedure which was first proposed and analysed by Symanzik \cite{Symanzik} in the 
context of scalar field theory,
 can be carried out loop order by loop order \cite{Osborn}.
For instance at one loop, one can evaluate the coincident limit of ${\cal{K}_{S}}$ by heat kernel methods,
This functional is a local functional of $K_{ab}$ and $g_{ab}$. The terms dependent on $K_{ab}$ can be reabsorbed into a redefinition 
of the kinetic terms while the terms dependent on $g_{ab}$ into a redefinition of the potential terms which cancels the divergence up to two loop\footnote{
In a renormalisable theory one can show that this procedure 
 preserves the Schr\"odinger form of the Hamiltonian, that is the hamiltonian operator is always at most quadratic in the functional derivatives $\hpi^{ab}$.
This follows from the fact that the renormalisation factor
which is constructed in terms of the boundary to boundary propagator  
contains at  most two time derivative.
In a theory of gravity this procedure make sense in dimension higher that $3$ only if one fixes the number 
of loops and allows oneself to fix a increasing number of renormalisation constant.}.

The last point we want to stress concerning the Hamiltonian equation is the obvious fact that this is a second order differential equation.
For comparison with the CFT equations, it is convenient to decompose the metric in terms of a Liouville field $\phi$ and a determinant one metric $\bga = e^{2\phi}\hat{\bga}$, $\det(\hat{\bga})=1$. 
We define $e^{-d\phi} \hat{P}^{a}_{b} \equiv \hpi^{a}_{b} -\frac{\delta^{a}_{b}}{d}\hpi$
the traceless derivative operator that acts on $\hat{\bga}$ only,
preserves its unimodularity 
and commute with $\delta/\delta \phi$.
The hamiltonian equation in this splitting can be written as 
\be\nonumber
{\cal H} =    \frac{\kappa^{2}}{d(d-1)} \left(\frac{\delta}{\delta \phi}\right)^{2} 
-{ \kappa^{2}}e^{-2d\phi}\hat{P}^{2} + 
e^{-2\phi} {R(\hat{\bga})}   - 2(d-1)\left(\hat{\Box} \phi + (d-2)(\hat{\nabla} \phi)^{2}\right) + \frac{d(d-1)}{ l^{2}}
\ee
where $\hat{P}^{2} = \hat{P}^{a}_{b}\hat{P}^{b}_{a}$.
This equation is a relativistic equation ( if one think that $\phi$ plays a role analogous to time) which controls the Liouville field dependence  of $\Psi(e^{2\phi}\hat{\bga})$.
Such dependence is in principle determined once  
{\it both} $\Psi(\hat{\bga})$ and $\frac{\delta}{\delta \phi}\Psi(e^{2\phi}\hat{\bga})|_{\phi=0} $ are given.
This means that one expect the existence of (highly non trivial) propagating kernels $K_{\phi}, \tilde{K}_{\phi}$ such that 
\be\label{Kernel}
\Psi(e^{2\phi}\hat{\bga}) = \int {\cal D} \hat{\bga}' \left(
K_{\phi}(\hat{\bga},\hat{\bga}') \Psi(\hat{\bga}') 
+\tilde{K}_{\phi}(\hat{\bga},\hat{\bga}')\frac{\delta \Psi}{\delta \phi}(\hat{\bga}') \right).
\ee
Taking $e^{2\phi} = \rho^{-2}$ with $\rho$ a spatial constant defines a 
one parameter hamiltonian subgroup or a preferred radial evolution.
In the case of deSitter this evolution is the usual cosmological evolution
where time appears as the rescaling of the spatial slices.

\subsection{CFT equations}

The analysis of  the CFT side is much simpler.
Since we work with pure gravity we are only interested by the equation 
satisfied by the CFT partition function $Z_{CFT}(\bga)$ which also depends like $\Psi$ on a d-dimensional metric $\bga$.
This CFT partition function is the generating functional of connected correlation function of the 
energy momentum tensor of the CFT since insertions of $T^{ab}$ can be obtained by derivative of $Z_{CFT}$ with respect to $\gamma_{ab}$.  
Such a CFT partition satisfies also two equations. The first one expresses the invariance under diffeomorphism 
\be\label{diffCFT}
 \nabla_{a} \hpi^{a}_{b} Z_{CFT}(\bga) =0,
\ee
and the second is the conformal Ward identity
\be\label{Ward}
\hpi_{x} Z_{CFT}(\bga)= \frac{1}{\sqrt{\gamma}}\left. \frac{\delta}{\delta \phi(x)} Z_{CFT}(e^{2\phi}\bga)\right|_{\phi=0}  = i A_{d}(x) Z_{CFT}(\bga)
\ee
where $A_{d}(x)$ is the anomaly. It is zero in odd dimensions, whereas in even dimension it can be expanded in terms of a basis of certain curvature invariants of dimension $2d$ which should satisfy the Wess-Zumino consistency condition
\cite{Anomaly}. In dimension $2$ and $4$ it is given by 
\bea
A_{2}(x) &=& \frac{c}{24\pi} R(x) \\
A_{4}(x) &=&  \frac{1}{16\pi}\left(  a E(x) -c W^{2}(x) + \alpha \Box R(x)\right)
\eea
where $c$ and $a$ are the two central charges.
$W^{2}$ is the square of the Weyl Tensor and 
$E$ is the Euler density. The term proportional to $\alpha$ is not an anomaly since it can be 
obtained from the variation of a local action $\int \sqrt{\bga} R^{2}$.
\bea
W^{2} &=& R_{abcd}^{2} - 2 R_{ab}^{2} +\frac13 R^{2},\\
E= \left(\frac12 \epsilon_{ab}^{ef}R_{ef cd}\right)^{2}&=& R_{abcd}^{2} - 4 R_{ab}^{2} +R^{2}.
\eea
One remark that when $a=c$ the 4 dimensional conformal anomaly simplifies and contains no square of the Riemann tensor. Interestingly this is exactly the anomaly that arise from the semi-classical AdS/CFT correspondence \cite{Skenderis1, oneloop}.

The conformal Ward identity is only  a first order equation which can be explicitly integrated out.
In two dimensions, the integrated Ward identity is given by 
\be\label{intanomaly}
Z_{CFT}(e^{2\phi} {\bga}) = e^{ \frac{i c}{24\pi} S_{L}(\phi, {\bga})} 
Z_{CFT}({\bga})
\ee
where the Liouville action is 
\be\label{liouville2}
S_{L}(\phi,\bga) = \int_{\Sigma}  \sqrt{\bga}\left( - \phi \Box \phi
+ \phi R(\gamma)\right).
\ee
In dimension 4, it is also possible to integrate out the anomaly. Quite remarkably, and this fact seems to have been unnoticed in the literature, in the gravitational case $a=c$ the integrated anomaly is also of 
the Liouville form\footnote{where the prefactor is now $\frac{c}{2\pi}$ and one need to chose  $\alpha= 4$} (\ref{intanomaly})
\be
S_{L}(\phi,\bga) = \int_{\Sigma} \sqrt{\bga}\left(\frac12\phi \Box_{4} \phi
+ \phi Q_{4}(\gamma) \right)
\ee
 where 
 $Q_{4}\equiv  \frac12\Box P +  P^{2} - P_{a}^{b}P_{b}^{a}$ 
and $\Box_{4} \equiv \Box^{2} + \nabla_{a}\left( 2Pg^{ab} - 4 P^{ab}\right)\nabla_{b}  $ with
$
P_{ab}
= \frac{1}{d-2}\left(R_{ab} -\frac{R}{2(d-1)}g_{ab }\right),\quad 
P  = \frac{R}{2(d-1)}
$.
These objects  satisfy
 the key relation
\be e^{4\phi} Q_{4}(e^{2\phi}\bga)= \Box_{4}\phi +  Q_{4}(\bga).\ee
The curvature $Q_{4}$ is part of a series of curvatures called Q-curvatures which are the object of intense
study in the mathematical literature \cite{Graham}.

\subsection{2d CFT}
It is important to remember that a general CFT of central charge $c$ in  dimension two is not only supposed to be invariant 
under infinitesimal diffeomorphisms as expressed by (\ref{diffCFT}) but also under large diffeomorphism, in other words it should also be modular invariant. 

In two dimensions, the conformal Ward identities are powerful enough to extract  
a wealth of information about energy-momentum tensor correlations.
As we have seen the
conformal anomaly can be integrated out in terms of the Liouville action (\ref{liouville2}).
If we pick local  coordinates $u,\tilde{u}$ we can write the metric as 
\be
ds^{2} = e^{2\phi}(du + \mu d\tilde{u})(d\tilde{u} +\tilde{\mu} du) = e^{2\phi} \hat{g}
\ee
where $\mu, \tilde{\mu}$ are the so-called Beltrami differentials.
And the Partition function decomposes  as
\be
Z(e^{2\phi} \hat{g}) = e^{\frac{ic}{24\pi} S_{L}(\phi,\hat{g})} Z(\hat{g}).
\ee
This decomposition is dependent on the choice of the local coordinate system and imply that the action of diffeomorphism on $Z(\hat{g})$ is anomalous.

What is quite remarkable and has been first shown by Verlinde \cite{verlinde}, in a beautiful work, 
is that it is possible  to chirally split the diffeomorphism anomaly if one add to the effective action the counterterm 
\be
S_{V}(\mu,\tilde{\mu}) =\int \frac{1}{1-\mu\tilde{\mu}}
\left(\partial_{u}\mu \partial_{\tilde{u}}\tilde{\mu} -\frac{\tilde{\mu}}2  (\partial_{u}\mu)^{2} -
\frac{\mu}2 (\partial_{\tilde{u}}\tilde{\mu})^{2}  \right)
\ee
Namely if one define $S(\phi,\mu,\tmu)= S_{L}(\phi,\hat{g}) + 
S_{V}(\mu,\tilde{\mu})$ and define
\be
Z(e^{2\phi} \hat{g}) \equiv e^{\frac{ic}{24\pi} S(\phi,\mu,\tmu)} Z(\mu,\tmu).
\ee
Then $Z(\mu,\tmu)$ satisfies the Virasoro Ward identities
\bea\label{Vir}
\left(\partial_{\tu} -\mu \partial_{u} -2\partial_{u}\mu \right)\frac{\delta Z}{\delta \mu(x)} (\mu,\tmu)
&=& \frac{ic}{12\pi} \partial_{u}^{3}\mu(x) \, Z(\mu,\tmu)\label{ward1}\\
\left(\partial_{u} -\tmu \partial_{\tu} -2\partial_{\tu}\tmu \right)\frac{\delta Z}{\delta \tmu(x)}(\mu,\tmu)
&=& \frac{ic}{12\pi} \partial_{u}^{3}\tmu(x) \, Z(\mu,\tmu).
\eea
Since derivative with respect to $\mu$ inserts expectation value of the holomorphic component of the 
energy momentum tensor: $Z(\mu,\tmu) = \bra e^{\frac{i}{\pi}\int T\mu}e^{\frac{i}{\pi}\int \tilde{T}\tilde{\mu}}\ket$, it is possible to show that these Virasoro ward identities are 
if fact equivalent to the usual Virasoro O.P.E.
\be
T(z)T(w)\sim \frac{c/2}{(z-w)^{4}} + \left( \frac{2}{(z-w)^{2}} + \frac{\partial_{w}}{(z-w)}\right)T(w).
\ee
The space of holomorphic conformal block is defined to be the space
of solutions of  (\ref{ward1}) and this implies that the partition function can be written as a sum of holomorphic times anti-holomorphic components
\be
Z(g) = e^{\frac{ic}{24\pi} S(\phi,\mu,\tmu)} \sum_{I,\tilde{I}} N^{I\tilde{I}} \chi_{I}(\mu) \chi_{\tilde{I}}(\tmu).
\ee
The only ingredient not fixed by the conformal symmetry is the value of $N^{I\tilde{I}} $ which should be such that  the CFT is modular invariant.
If we assume that the spectra of the CFT is discrete and then $N^{I\tilde{I}}$ should be  positive integers. A general solution of the Ward identity can be obtained form linear  combination of such irreducible CFTs.

\section{Puzzles resolution and AdS/CFT Dictionary}
\label{sec2}
Now that we understand the nature of both sides of the AdS/CFT correspondence in more details, we can present a resolution of the three puzzles presented earlier.
 The key puzzle to address first, and which naturally leads to the resolution of the other two, is the question about background independence.
 The gravity AdS-matrix is defined by 
 \be
\Psi_{{\Sigma}}(\bga) =\int_{g|_{\partial M}= \bga}\!\!\!\!\!\!\!\mathrm{D} g \,\,e^{i S_{M}(g)}  
\ee
The key puzzling point  is that such a partition function depends on a topological manifold $M$ with boundary and a choice of a boundary metric $\bga$.
However this amplitude do not depend on some auxiliary asymptotically AdS spacetime and corresponding bulk metric. 
We cannot tell for instance where is the boundary located in some asymptotic AdS space.  But if so where is asymptotic infinity?

 In fact, because we are dealing with quantum gravity the functional $\Psi$ {\it represents} the quantum spacetime. 
This quantum spacetime is however represented only through its dependence on the boundary metric $\bga$. In a classical spacetime this boundary metric is the metric induced on a radial slice and changing the 
slice will change the induced boundary metric.
In quantum gravity we just have to reverse this point of view:  varying the  metric $\gamma$ amounts to move the slice inside the spacetime represented by $\Psi$. The knowledge of $\Psi$ for all possible values of $\gamma$ allows us in principle to reconstruct the semi-classical spacetime it corresponds to.
At the classical level we have seen that the metric induced on the radial slice $\Sigma_{\rho}$ is given near infinity by 
$\rho^{-2}\gamma$ where $\gamma $ is a representative of the conformal class of the metric induced on asymptotic infinity.
The rescaling in $\rho$ amounts to move the slice towards infinity.
The natural proposal is therefore that the asymptotic property of  the spacetime  encoded by $\Psi$ 
is given by the behavior of the wave functional $\Psi(\rho^{-2}\gamma) $ in the limit when $\rho \to 0$.
So asymptotic infinity in this context is not a specific locus in a background metric but 
comes from the asymptotic behavior of the functional $\Psi(\bga)$ under infinite rescaling.
This is a simple but very important point that allows us to talk about asymptotically AdS spacetime in quantum gravity without having to introduce a background spacetime.

\subsection{The holographic side of Wheeler-deWitt equation}

At the classical level the properties of asymptotic infinity follows {\it dynamically} from the Einstein equation once one assumes the existence of a conformal boundary. 
The idea is simply to follow the same reasoning at the quantum level by looking at  the asymptotic behavior of a solution of the radial WdW equation.
The main property we want to show in this section is the following:

Given $\Psi(\gamma)$ an arbitrary solution of the radial WDW equation (\ref{WdW}) then its asymptotic behavior when 
$\rho \to 0$ is given by
\be\label{asym}
\Psi_{\Sigma}\left(\frac{\gamma}{\rho^{2}}\right) \sim
 e^{\frac{i}{\kappa} S^{(d)}\left(\frac{\gamma}{\rho^{2}}\right)} Z_{+}\left(\frac{\gamma}{\rho^{2}}\right) +
e^{-\frac{i}{\kappa}S^{(d)}\left(\frac{\gamma}{\rho^{2}}\right)} Z_{-}\left(\frac{\gamma}{\rho^{2}}\right)
\ee
where $S^{(d)}(\bga)$ is an explicit {\it local}   action, 
containing terms of dimension at most $d$. This action is {\it universal } in the sense that 
it is the same for all radial states  $\Psi$.
Moreover  $Z_{\pm}(\gamma) $ are a pair of CFT's solution of the conformal Ward identity (\ref{Ward}).
\be
\frac{1}{\sqrt{\gamma}}\left. \frac{\delta}{\delta \phi(x)} Z_{\pm}(e^{2\phi}\bga)\right|_{\phi=0}  = \pm i A_{d}(x) Z_{\pm}(\bga)
\ee

This means that we can easily extract the $\rho$ dependence of $Z_{\pm}$ and get
\be\label{asym2}
\Psi_{\Sigma}\left(\frac{\gamma}{\rho^{2}}\right) \sim
 e^{\frac{i}{\kappa} \tilde{S}_{\rho}^{(d)}(\bga) }  Z_{+}(\gamma) +
e^{-\frac{i}{\kappa}\tilde{S}_{\rho}^{(d)}(\bga) } Z_{-}(\gamma).
\ee
where 
$$\tilde{S}_{\rho}^{(d)}(\bga) = S^{(d)}\left(\frac{\gamma}{\rho^{2}}\right) +
 \kappa \ln\rho \int_{\Sigma} \sqrt{\bga} A_{d}.$$
The conformal Ward identity satisfied by $Z_{\pm}$ is a left over or more precisely an holographic in-print of the radial WdW equation 
onto asymptotic infinity. Moreover from (\ref{asym2}) and the previous discussion it is clear that the CFT's $Z_{\pm}$ can be thought  simply as initial value data 
for the radial wave function $\Psi$. The only unusual feature being that this initial slice is at infinity instead of being in the bulk.
The expansion  (\ref{asym2}) is also reminiscent of a Born-Oppenheimer expansion were infinity is treated as the ``heavy'' component and 
possesses 
a rapidly oscillating but  localized phase factor, whereas the CFT is the leftover ``light'' component.

The local action can be computed in terms of a loop expansion, the order zero term is, as we will see, given\footnote{We have denoted
 $ P_{a}^{b} \equiv \frac{1}{(d-2)} \left(R_{a}^{b} - \frac{R}{2(d-1)} \delta_{a}^{b}\right) $ so that the 
 third lagrangian  is $L_{2}=\frac{1}{(d-2)^{2}}\left(R_{a}^{b}R_{b}^{a}- \frac{d}{ 4(d-1)}
 R^{2}\right)$ }
\be\label{S_{l}}
S^{(d)}(\bga) = 
\frac{d-1}{\ell} \int_{\Sigma} \sqrt{\bga} - 
 \frac{\epsilon \ell}{2(d-2)} \int_{\Sigma}\sqrt{\bga} R(\bga) +
 \frac{\ell^{3}}{2(d-4)}\int_{\Sigma}\sqrt{\bga} \left( P_{a}^{b}P_{b}^{a}- PP\right)+\cdots 
 \ee
 This expression is exactly the same as the counterterm action appearing  in the 
holographic renormalisation group analysis \cite{ Skenderis2, Myers}.
 In dimension $d=2$ only the first term of (\ref{S_{l}}) is relevant in the limit 
 $\rho \to 0$;
 in dimension $d=3,4$ the first two terms contribute; in dim $5,6$ the first three etc...
 One also notice that in even dimension, the integrated anomaly is exactly given by the 
 residue of the pole that arises for $d=2n$ in the local expansion of $S^{(d)}$.

  The inclusion of loop corrections renormalises the coefficients appearing in this local expansion.
  This renormalisation can be explicitly computed at one loop and we will see later
  that in $d=2$ it leads to a finite renormalisation of the  central charge.
  Since only a finite number of terms are relevant in order to get the dominant asymptotic, it seems that 
  the non renormalisability of gravity is not registered by the asymptotic expansion (\ref{asym}), this is clearly something that deserves further study and a deeper understanding.
  
To see why a solution of the radial WdW equation has  the asymptotic behavior (\ref{asym}) 
one first perform the change of variable $\bga_{ab} \to \rho^{-2}\bga_{ab} $; $ \hpi^{a}_{b} \to \rho^{-d} \frac{2}{\sqrt{\bga}}\bga^{ac}\frac{\delta}{\delta \bga_{cb}}\equiv
  \rho^{-d} \hpi^{a}_{b}$ in the hamiltonian constraints.
   Thus $\Psi(\rho^{-2}\bga)\equiv \Psi_{\rho}(\bga) $ is a solution of  ${\cal H}_{\rho} \Psi_{\rho}(\bga) =0$ with
 \be
 {\cal H}_{\rho} =   -{\epsilon \kappa^{2}\rho^{2d}} \,
\hpi\cdot \hpi 
+\epsilon \frac{d(d-1)}{ l^{2}}  + 
 \rho^{2} {R(\bga)}  
\ee
 where we have denoted $\hpi\cdot \hpi \equiv \hpi_{a}^{b}\hpi^{a}_{b} -\frac{\hpi^{2}}{d-1} $. We now look here only at the radial equation 
 for AdS ($\epsilon$ =+1). The dS case is similar.
 
 One sees that in the limit $\rho \to 0$ the curvature term becomes irrelevant.
 and the solution of the resulting equation, where it is neglected, is easily found There are   two such solutions given by 
$$\Psi^{(0)}_{\rho}(\gamma) =\exp\left(\pm\frac{i}{ \kappa \rho^{d}}
\frac{d-1}{\ell}
\int_{\Sigma} \sqrt{\bga}\right)$$
a linear combination of which represents  the dominant term in the asymptotic expansion of $\Psi_{\rho}$.
They satisfies 
$$
 { \kappa^{2}\rho^{2d}} \,
\hpi\cdot \hpi \Psi_{\rho}^{(0)}(\bga)=
\frac{d(d-1)}{ l^{2}} \Psi_{\rho}^{(0)}(\bga).
$$   
  We can now expand around the state associated with
  the sign + (the expansion around the other state is similar) : $\Psi =\Psi^{(0)}\Psi^{(1)}$.
  This expansion is easily carried out if one
  use the fact that 
  $  \rho^{d} \hpi^{a}_{b}\Psi =\Psi^{(0)}\left(i\frac{d-1}{\kappa \ell}\delta^{a}_{b}+\rho^{d}\hpi^{a}_{b}\right)\Psi^{(1)}$.
  By construction the argument of $\Psi^{(1)}$  is subdominant ($O(\rho^{2})$)compare to the one of $\Psi^{(0)}$, thus $(\Psi^{(1)})^{-1}\left(\rho^{d}\hpi^{a}_{b} \Psi^{(1)}\right)$ is of order at least $\rho^{2}$.
 In this expansion the hamiltonian constraint becomes 
 \be
 i \frac{2 \kappa}{\ell}  \rho^{d} \hpi \Psi^{(1)}+ 
 \rho^{2} {R(\bga)}  \Psi^{(1)}
 - {\kappa^{2}\rho^{2d}} \, \hpi\circ \hpi  \Psi^{(1)} =0.
 \ee
  
  In the case $d=2$ and in the limit $\rho \to 0$ 
  the last term of the identity is negligible since $\hpi^{a}_{b} \Psi^{(1)}$ is $O(1)$ and the equation reduces, as promised,  to the Ward identity
  \be
 \hpi \Psi^{(1)}= 
i \frac{\ell}{2\kappa} {R(\bga)}  \Psi^{(1)}= i \frac{c}{24\pi} {R(\bga)}  \Psi^{(1)},\quad \mathrm{with} \quad 
c= \frac{3 \ell}{2G}.
 \ee
To go to the next order, we continue the expansion $\Psi= 
\Psi^{(0)}\Psi^{(1)}\Psi^{(2)}$ with 
$$ \Psi^{(1)}= e^{- i  \frac{\ell}{2\kappa(d-2)} \int_{\Sigma}\sqrt{\bga} R(\bga)}$$
the equation for $\Psi^{(2)}$ is then\footnote{If one neglects some operator ordering ambiguity that should be taken into account 
as a one loop renormalisation effect}
\beq\nonumber
 i\frac{2\kappa}{\ell} \hpi\Psi^{(2)} &+&  \frac{\rho^{4-d}}{(d-2)^{2}}\left(R_{a}^{b}R_{b}^{a}- \frac{d  R^{2}}{ 4(d-1)}\right)
 \Psi^{(2)}\\
&+&  \frac{i 2\kappa \ell \rho^{2}}{d-2} \left(R_{a}^{b}\hpi_{b}^{a} - \frac{2d-3}{2(d-1)} R \hpi\right) \Psi^{(2)}\nonumber
- {\kappa^{2}\rho^{d}} \, \hpi\circ \hpi  \Psi^{(2)} =0
\eeq
In dimension $4$ and in the limit $\rho \to 0$ it becomes
again the Ward identity
\be
\hpi \Psi^{(2)} =  i \frac{\ell}{8\kappa}\left(R_{a}^{b}R_{b}^{a}- \frac{  R^{2}}{3}\right)
 \Psi^{(2)}.
\ee
This shows the main statement of this section.

\subsection{Solving the Hamilton-Jacobi equation}

The result we have just presented is very close in spirit to the seminal work of  deBoer and Verlinde's \cite{deBoer}. This suggests
 a much simpler and more systematic derivation of the local action  that enters the asymptotic expansion at the semi-classical level.
This derivation essentially amounts to solve the Hamilton-Jacobi equation in terms of a local expansion.
Such an expansion have been studied already in \cite{HJ1} and developed to a much deeper extent by Skenderis et Papadimitriou
in \cite{HJ2}. Here we propose an even simpler derivation of their result and gives a closed recursive equation which follows directly from the Hamilton-Jacobi equation.

We define functional derivative operators 
\be
\hpi^{ab}_{x}\equiv \frac{2}{\sqrt{\gamma}}\frac{\delta}{\delta g_{ab}(x)},\quad  \hpi^{a}_{b} \equiv \gamma^{ac}\hpi_{cb}, \quad \hpi \equiv \hpi^{c}_{c}, 
\quad
\hK^{a}_{ b}\equiv \hpi^{a}_{b} - \delta_{b}^{a} \frac{\hpi}{d-1},\quad
\Pi_{a}^{b} = \hK^{a}_{ b} - \delta_{a}^{b} K
\ee
The Hamiltonian constraint (for the AdS case) is given by 
\be
\frac{\cal H}{\sqrt{|\gamma|}} =  -{\kappa^{2}}\left(\hpi^{a}_{b}\hpi^{b}_{a} -\frac{\hpi^{2}}{d-1}\right) +  \left( {R(g)}  + \frac{d(d-1)}{ l^{2}}\right)
\ee
Looking for a state $\Psi = e^{\frac{i}{\kappa} \tilde{S}(g)}$ solution of the Hamiltonian constraint and neglecting all quantum correction proportional to $\kappa$ we get the Hamilton-Jacobi 
equation. For future convenience we write down this equation in terms of the $\hK$ operator instead of the canonical momenta operator. One of the reason is that when we act with this operator on $S$ we obtain the extrinsic curvature which has a clearer geometrical meaning.
It is also convenient to introduce the short hand notation
\be
(K\circ K)_{\mu\nu} \equiv \bga^{\alpha\beta} 
(K_{\mu \alpha}K_{\nu\beta}-K_{\mu \nu}K_{ \alpha \beta}),\quad 
(K\circ K) \equiv \bga^{\mu\nu}(K\circ K)_{\mu\nu}
\ee
The corresponding Hamilton-Jacobi equation reads 
\be
\left(\hK(\tilde{S}) \circ \hK(\tilde{S})\right)(\bga)+   {R(\bga)}  + \frac{d(d-1)}{ l^{2}} =0.
\ee
We look for an expansion of $S(\bga)$ in terms of functional over the metric having  fixed conformal dimension, 
that is 
\be
\tilde{S}(\bga) = \sum_{n=0}^{\infty}S_{n}(\bga), \,\,\mathrm{with}\,\,\,\,\,
S_{n}\left({\rho^{-2}}{\bga}\right)
=\rho^{-d + 2n} S_{n}(\bga).
\ee
This expansion can therefore  be thought as a Taylor expansion in the parameter $\rho$.
Lets start to solve the equation at order $0$.
From
\be
\hK_{a}^{b}\left(\int \sqrt{\bga}\right) = \frac{1}{(1-d)} \delta_{a}^{b},\quad (\delta\circ \delta) = d(1-d),\quad (\delta\circ K) = (1-d) K
\ee
one easily sees that 
\be
S_{0}(\bga)= \pm \frac{(d-1)}{\ell}\int_{\Sigma} \sqrt{\bga}
\ee
is the solution at  order $0$.
In order to match the classical expansion we choose the sign $+$ that match the classical analysis that is $ \ell \hat{K}_{a}^{b}(\Psi) =  \delta_{a}^{b} \Psi$. The other sector can be obtained by changing the sign of $\ell$.
We can now use this solution to start a local expansion of the Hamilton-Jacobi equation,
$$\tilde{S}(\bga)= S_{0}(\bga) + {S}(\bga)$$ where $\hat{S}$ starts at order one.
The equation now reads 
\be
\frac{2}{\ell}\hat{\delta}^{D}_{x}S (\bga) = \left(\hK_{x}(S)\circ\hK_{x}(S)\right)(\bga) + R(\bga(x)) 
\ee
Where we have introduced the operator $\hat{\delta}^{D}_{x}$ that generates local conformal rescaling namely
\be
 \frac{2}{\sqrt{\gamma}}g^{ab}\frac{\delta}{\delta g_{ab}(x)} \equiv \hat{\delta}^{D}_{x} F(\bga) = \left.\frac1{\sqrt{\bga}}\frac{\delta F(e^{2\phi} \bga)}{\delta \phi(x)}\right|_{\phi=0}
\ee
If one integrate this operator over $\Sigma $ one obtain the conformal dimension operator, which can in turn be written as a differential operator $-\rho\partial_{\rho}$. 
Using this we can now get, quite remarkably, a closed equation for ${S}$ which is the result we were looking for
\be
2\partial_{r} S(e^{\frac{2r}{\ell}}\gamma) = \int_{\Sigma}\sqrt{\bga(x)} \left(\hK_{x}(S)\circ\hK_{x}(S)\right)(e^{\frac{2r}{\ell}}\bga) + e^{(d-2)\frac{2r}{\ell}}\int_{\Sigma}\sqrt{\bga}R(\bga).
\ee
We can write this equation in terms of the expansion coefficients
\beq
S_{1}(\bga) &=& \frac{\ell}{2(d-2)} \int_{\Sigma}\sqrt{\bga} R(\bga)\\
2(d-2n) S_{n}(\bga) &=& \ell \sum_{m=1}^{n-1} \int_{\Sigma}\sqrt{\bga} \left(\hK_{x}(S_{m})\circ\hK_{x}(S_{n-m})\right)(\bga)\label{SS}
\eeq
We can recursively solve these equations, the procedure is straightforward but increasingly tedious. It is however quite  simpler than the original way of computing the counterterm action
\cite{Skenderis1, Skenderis2}.
Using the results for $S_{1}$ we can compute the extrinsic curvature at first order and putting it back in (\ref{SS}) compute $S_{2}$, denoting $K_{(n)}{}_{a}^{b} \equiv\hK_{a}^{b}(S_{n})$ one obtains

\beq
K_{(1)}{}_{a}^{b} &=& -\frac{\ell}{(d-2)} \left(R_{a}^{b} - \frac{R}{2(d-1)} \delta_{a}^{b}\right) \equiv
- {\ell} P_{a}^{b} \\
S_{2} &=&  \frac{\ell^{3}}{2(d-4)}\int_{\Sigma}\sqrt{\bga} \left( P_{a}^{b}P_{b}^{a}- PP\right).
\eeq

\subsection{AdS/CFT dictionary}
\label{seckey}
The   asymptotic expansion (\ref{asym}) is the key formula 
expressing a deep relationship between bulk quantum gravity and boundary CFT. It  shows that a generic radial state solution of the radial Wheeler-deWitt equation  generically corresponds to {\it two } different CFTs.
And that these CFTs can be understood as initial value data, on 
a slice at infinity, for the wave function solution of radial WdW.

The fact that there are two CFTs is in sharp contrast with the usual interpretation of AdS/CFT where one usually assumes the existence of only one CFT associated with 
 quantum gravity in the bulk.
This surprising feature follows immediately from the fact that the WdW equation is second order whereas the Ward identity is first order so that we need two independent pieces of initial data to specify a solution of WdW equation.
Note that we have emphasized here the AdS case, but as we have already said all the derivations and conclusions are true also in the deSitter case. In the deSitter case, since one can have two asymptotic region, it is customary to associate two CFTs to a gravity state \cite{dS}, what we are emphasizing is that this conclusion should hold generically in AdS too.

Another surprising, and disturbing at first sight,  feature of our derivation is the fact that contrarily to the classical case we haven't assume at any moment that the state $\Psi$ corresponds to a spacetime which admits an asymptotic infinity.
In fact, the proof given in the previous section is valid for {\it any} solution of the quantum gravity equation of motion.
Among these there should be  ones which correspond semi-classically to AdS compact spacetime.
This means that the AdS/CFT correspondence can be in principle extended even to spacetime having no classical asymptotic infinity
as long as we accept having to deal with pairs of CFTs.
From the quantum point of view there is nothing deeply surprising about it since we know that a wave function do carry some information about classically forbidden region\footnote{
There is even in the mathematical literature an important result that points in the same direction: In the context of 3d classical gravity 
the problem of classification of all global hyperbolic Lorentzian AdS spacetimes with {\it compact } slice has been solved, single-handedly, in a remarkable work  by G. Mess \cite{Mess}.
Remarkably, one of the key ingredient of the constructive proof given by Mess is to show that a 3d geometry can be mapped to a pair of points on Teichmuller space, that is to a pair of conformal 2d geometries. These conformal geometries lives in fact  on the infinity of the universal cover of the compact spacetime under consideration.
Since the spacetime is compact this infinity is fictitious from the point of view of the initial spacetime geometry 
even if it the most efficient way to describe the 3d geometrical information.}.

Now this lead immediately to a natural question: Which quantum states $\Psi$ corresponds to spacetimes having an asymptotic infinity?
If one look closer at the structure of the asymptotic one can see that the two terms that arises  in the asymptotic expansion
are distinguished by the value of the extrinsic curvature (\ref{piK})
\be
\hat{K}_{a}^{b} \Psi = \frac{\kappa}{i}\left(\hpi_{a}^{b} - \delta_{a}^{b}\frac{\hpi}{d-1}\right)\Psi
\sim 
(\delta_{a}^{b} + \cdots) e^{iS^{(d)}(\gamma)} Z_{+}(\gamma)
+( -\delta_{a}^{b} + \cdots) e^{-iS^{(d)}(\gamma)} Z_{-}(\gamma)
\ee
where $\cdots$ stands for terms of order $O(\rho^{2})$.
The branch associated to $Z_{+}$ has a positive extrinsic curvature proportional to the identity
when going to infinity (represented by the limit $\gamma\sim \rho^{-2}\gamma$, $\rho \to 0$) which is what is expected in a 
classical spacetime with asymptotic infinity.
The branch associated with  $Z_{-}$ has has a negative extrinsic curvature 
when going to infinity. There is no classical interpretation of this branch.
The only possible interpretation is to imagine that one has  a slicing of spacetime where infinity is reached from the outside instead of the inside.
In the Schr\"odinger picture, the branch $Z_{+}$ correspond to a wave outgoing at infinity 
and the branch $Z_{-}$ to a wave in-going from infinity.
Thus if $Z_{-}\neq 0$ the wave asymptotic is similar to a standing wave for a particle in a confining potential.
This analogy and the extrinsic curvature computation tell us that a radial state can correspond to a
semi-classical spacetime with asymptotic infinity if we have that $Z_{-}=0$. 
From now on we will assume that an quantum AdS spacetime with asymptotic infinity is a solution of radial WdW equation
where $Z_{-}=0$. This is then consistent with the usual picture. However, at this stage this is only  a plausible  interpretation 
 that $Z_{-}=0$ amounts to having an asymptotic infinity. More work is clearly needed in order to show this  in more detail and understand the deeper role of the pairs of CFT's\footnote{
 Note that if we where working in the context of Euclidian AdS/CFT then the behavior of a radial WdW solution is 
 \be
 \Psi_{\Sigma}\left(\frac{\gamma}{\rho^{2}}\right) \sim
 e^{-\frac{1}{\kappa} \tilde{S}_{\rho}^{(d)}(\bga) }  Z_{+}(\gamma) +
e^{+\frac{1}{\kappa}\tilde{S}_{\rho}^{(d)}(\bga) } Z_{-}(\gamma).
 \ee
where $\tilde{S}_{\rho}^{(d)}(\bga)$ is a {\it positive} functional near $\rho=0$
and $\frac{1}{\sqrt{\gamma}}\left. \frac{\delta}{\delta \phi(x)} Z_{\pm}(e^{2\phi}\bga)\right|_{\phi=0}  = \mp A_{d}(x) Z_{\pm}(\bga)$.
The universal prefactor associated with $Z_{+}$ is  exponentially subdominant in the limit compare to the factor associated with $Z_{-}$.
The semi-classical evaluation of the Euclidean path integral corresponds to the sector 
$Z_{-}=0$ for which $\Psi_{\Sigma}$ is normalisable.
}.

Another key point which follows from (\ref{asym}) is the fact that the correspondence between quantum gravity and Conformal Field theory is not one to one, that is there is not one CFT  which is equivalent to the theory of quantum gravity, even if we restrict to $Z_{-}=0$.
The correspondence is indeed one to many.  More precisely, since the CFT is understood as some initial value data for the quantum gravity radial states, it defines a radial state. 
In fact as we will see in the next section the correspondence should be such that for {\it any} boundary CFT there is 
a corresponding quantum gravity radial state solution of WdW equation as long as the central charge match.
This will be argued for the case of general dimension and proven in the case $d=2$.
For instance, in the case $d=2$ what this means is that the space of  2d CFTs with fixed central charge is {\it isomorphic} to the 
space of radial quantum gravity states.

This is not at odd with the usual interpretation in which a drastic change in the bulk (addition of branes or different internal boundary conditions) can be sometimes taken into account in the CFT by adding  a 
relevant or irrelevant operator which modify the hamiltonian of the original CFT.
However, the picture here is even more dramatic than that (in the case of pure gravity and in the absence of SUSY\footnote{For instance, the set all possible CFT in dimension $4$, which have $N=4$ SUSY and a fixed anomaly (which determine the gauge group)  consist  of only one candidate \cite{WittenCFT}. }), because scanning the space of CFT means changing the Hamiltonian with a given field content but it also means that we can change the field content dramatically as long as we preserve the central charge. It can also mean that we scan different CFT having quite different gauge symmetries. 

To some extent  AdS/CFT correspondence in the context of pure gravity is
reminiscent of  what happens in the quantum Hall effect
In this case the bulk system is made up of interacting electrons
confined to a plane. Now, given a choice of filling fraction, magnetic field, etc... that fix a vacuum state in the bulk (e.g Laughlin state) there exists a boundary description of this system in terms of a chiral CFT. The chiral boundary theory depends heavily on the chosen bulk vacuum  state.
The difference here  is that what is fixed in the bulk is a radial state and not an usual hamiltonian vacuum state.
The physical meaning and interpretation of this radial state is not entirely clear, despite some attempts \cite{Oeckl, Giddings}.
Bringing light on this issue seems one of the key thing to develop in order to grasp the deeper meaning of AdS/CFT in the context of quantum gravity.

What we can say is that a radial state represents at the semi-classical level a spacetime or a superposition of spacetimes with 
fixed boundary topology. So it is somehow a definition of what we would call a quantum spacetime.
At the quantum level it is not known precisely  how this radial state is related to usual hamiltonian states of gravity,
but what one can expect is that such a radial states are in one to one correspondence with density matrix, even the precise form of the isomorphism needs to be unraveled.

Since it is clear that there are many CFT's given one quantum gravity theory,
 we are faced with the puzzle of understanding what does it means to find {\it the} CFT dual to a given theory of quantum gravity?
For instance in \cite{Witten3d} Witten made a proposal  for {\it the}  CFT  dual to 3d gravity.
How can we prove or disprove such a statement even in principle, if the correspondence is not one to one and assuming that we have a 
bulk definition of quantum gravity?

From the previous discussion we know that a CFT cannot define the full quantum gravity theory 
but at most a radial state of quantum gravity, thus what really make sense is to identify a particular radial state of quantum gravity 
and then construct {\it  the} CFT associated to this preferred ``vacuum'' radial state.
The question is then which radial state should one chose?

If we first think about this question in the Euclidean context, there is a preferred notion of such a radial state (see \cite{WM} for a 3d example).
Let us recall that given a topological manifold $M$ with boundary $\Sigma$ and assuming that we have a definition 
of bulk quantum gravity (i-e that we can make sense of the path integral ) we can associate to such 
a topological manifold a unique radial state given by
\be
\Psi_{M}(\bga) = 
\int_{g|_{\Sigma}= [\bga]}\!\!\!\!\!\!\!\mathrm{D} g \,\,e^{i S_{M}(g)}  
\ee
where $[\bga] $ denotes a diffeomorphism equivalence class of boundary metric.
More precisely Lets consider $D_{\Sigma}\equiv  Diff_{\Sigma}/Diff_{M}(\Sigma)$ the space of diffeomorphisms of $\Sigma$ 
which do not extend to  diffeomorphism of $M$ ($D_{\Sigma}$ only contains diffeomorphism not connected to the identity). 
Then
\be
\Psi_{M}(\bga) = \sum_{f\in D_{\Sigma}}
\int_{g|_{\Sigma}= f^{*}\bga }\!\!\!\!\!\!\!\mathrm{D} g \,\,e^{i S_{M}(g)} .
\ee
The summation over non trivial boundary diffeomorphisms ensure the modular invariance of the boundary CFT.

This prescription assigns modular invariant states to spacetimes. It is then natural to 
propose that the preferred radial state one should study corresponds to the simplest manifold given a boundary topology.
For instance, if  $\Sigma$ is a d-dimensional sphere one should  take as $M$ the $d+1$ dimensional ball. If  $\Sigma$ is a d-dimensional torus, we should take as $M$ the handlebody or plain torus which is the simplest topological manifold 
having $\Sigma$ as a boundary.
This defines a unique radial state of quantum gravity and a corresponding CFT
that should be identified.
If $d=2$ and $\Sigma$ is a genus $g$ surface we can still define the notion of a handlebody $H_{\Sigma}$ with boundary $\Sigma$
and the corresponding state $\Psi_{H_{\Sigma}}$ should allow us to identify  {\it the} genus $g$  CFT partition function.

The prescription for the Lorentzian case which is really the case of interest is a bit more difficult to identify precisely.
The reason being that a radial state depends only on the timelike part of the boundary.
However, in general one needs to specify what type of data one should use in the spacelike part of the boundary.
For instance, if we have an AdS cylinder, it is natural to put as initial and final state the vaccua state of the 
ADM hamiltonian which will be  identified as some time translation operator on the boundary.
Thus the boundary amplitude, will not really be a partition function but a transition amplitude from initial to final CFT state.

In order to identify the CFT we would like to have access to a partition function instead. Moreover, the 
radial gravity state that one should use to identify the CFT should at least contain Black-Holes,
 so it is natural to consider $M$ to be a manifold of topology the one of an eternal Black Hole solution in AdS.
Such a space has trivial topology in the Bulk and  
possess two different asymptotic region which are both isometric to  timelike cylinder on which one should fix  the  asymptotic conformal metric. 
There is an ambiguity to resolve onto how one should identify the two metrics associated with different timelike cylinders. The simplest and most natural prescription is to identify these two metrics at late and early time so that the asymptotic boundary is effectively a torus. 
This leads to a proposal (which should be tested) for a choice of a radial state of gravity 
which is needed if one want to be able to identify, 
even in principle, what is meant by  {\it the} CFT associated with gravity.

There is an important caveat to keep in mind in this derivation: as we have seen, from a radial state of the type $\Psi_{H_{\Sigma}}$ we can extract the CFT partition function $Z_{+}(\bga)$ which satisfies all the axioms that a CFT partition function should satisfy.
However, it is not clear a priori that such a CFT partition function is irreducible. 
That is we haven't shown that such a CFT partition function cannot be written a  linear sum of different CFTs: 
$Z_{+}(\bga) = \sum_{i} a_{i} Z_{i}(\bga)$ where $Z_{i}$ are individual  CFTs associated with a  conformally invariant Hamiltonian $H_{i}$.

\section{Bulk reconstructing Kernel}
\label{bulkrec}

In the previous section we have seen that given a radial quantum gravity state $\Psi$ we can extract from it 
a CFT $Z_{+}$ (we restrict to gravity states determined by one CFT) by looking at its asymptotic behavior.
 The key question now is can we  reconstruct the bulk geometry, 
hence $\Psi$, from the knowledge of the Boundary CFT?

 The answer to this question is clearly positive: First, as shown in section \ref{grav}, since $\Phi$ is solution of an Hamiltonian equation we can reconstruct it from its knowledge on a given initial slice (see (\ref{Kernel})); moreover 
 we can interpret  $Z_{+}$ as the specification of $\Psi$ on a slice situated at infinity.
One expect, therefore, the existence of a reconstructing kernel which allows us to ``evolve'' 
 $\Psi$ from infinity to the the interior. This means that there exist a Kernel $K(\gamma,\gamma')$ which is 
 state independent and should  intertwine the action of the radial hamiltonian constraint with the action 
 of the  Ward identity operator. That is $K$ should be such that 
 \be\label{BR}
 \Psi_{Z}(\bga) = \int D\bga' K(\bga,\bga')  Z(\gamma'),\quad \quad  {\cal H} \Psi_{Z} =0
\quad\mathrm{and} \quad  \Psi_{Z}(\rho^{-2}\bga) \sim_{\rho=0} e^{i \tilde{S}^{(d)}(\rho^{-2}\bga)}Z(\bga),
 \ee
 given any solution $Z$ of the conformal ward identity.
 
 It is often  said or assumed \cite{deBoer} that this bulk reconstruction is equivalent to a renormalisation group analysis
 hence the name ``holographic renormalisation''. However in the light of our analysis we don't feel that this is a very accurate description:
 It is true that a change of scale in the boundary (going from small to big scale) amounts to probing more and more the interior of the bulk \cite{WittenLenny}. 
 In fact this comes from the original picture that a rescaling of $\bga$ as an argument of $\Psi$ can be interpreted as moving 
 the slicing which defines $\bga$ radially.
 Moreover, (this follows from (\ref{BR}) and we will this more explicitly soon)
  the kernel   $K$  is picked  around $\bga=\bga'$ so a a bulk metric rescaling amounts to a boundary metric 
  rescaling.  However, it is incorrect to think that the equation governing the radial evolution is just a renormalisation group equation (which is first order). The renormalisation group equation (or conformal Ward identity) governs finite rescaling of the boundary CFT. In order to probe the bulk we need to achieve infinite rescaling and the corresponding equation is the full WdW equation which is second order.
 In this sense the radial Wheeler de Witt equation can be thought as an extension or a completion of the renormalisation group equation 
 beyond its usual range of validity.  And the Kernel $K$ is needed in order to convert boundary rescaling into bulk radial motion.

\subsection{Reconstructing Kernel in 2+1 dimensions}

We now present an explicit  reconstruction formula for the wave functional in terms of the boundary CFT
 in the case of three dimensional gravity.
In order to do so it is convenient to label the 2-dimensional geometries by frame fields   $e^{\pm}=e^{\pm}_{\mu}dx^{\mu}$ and 
$E^{\pm}$  such that $$ds^{2}= e^{+}e^{-},\quad ds'^{2}= E^{+}E^{-}.$$ 
We also denote $$e \equiv e^{+}\wedge e^{-}$$ the 2d volume two form\footnote{The normalisation is such that $e =2\sqrt{\bga}$.}.

The main claim of this paper is that { if  $Z_{c}(E)$ is a solution of the 2d Ward 
identity then the wave functional $\Psi(e^{\pm})$:
\be\label{main}
\Psi\left(e^{\pm}\right)=
\exp\left( \frac{ik}{4\pi}\int_{\Sigma} e \right)
\int DE \,
\exp\left( 
-\frac{ik}{2\pi} \int_{\Sigma} (E^{+}-e^{+})\wedge (E^{-}- e^{-})\right)
Z_{c}(E)
\ee
 is a solution of the three dimensional Wheeler-deWitt equation.}
 Here $k \equiv \frac{\ell}{4G}$ and the relationship between $k$ and $c$ is 
\be
c=1+6k.
\ee
The functional measure is given in term of the reparametrisation and Lorentz gauge invariant distance on the space of frame
\be\label{dist}
(\delta E, \delta E) 
= \int \delta E^{+}\wedge \delta E^{-}\quad 
\int D(\delta E) \,\, e^{i \frac{k}{2\pi}(\delta E, \delta E)}\equiv 1.
\ee
Before giving a proof of this statement  lets us consider the behavior of such a state $\Psi$
under a rescaling of the metric:
\be\nonumber
\Psi\left(\frac{e^{\pm}}{\rho}\right)=
\exp\left(  \frac{ik}{4\pi \rho^{2}}\int_{\Sigma} e \right)
\int D\left(\frac{E}{\rho}\right) \,
\exp\left( 
-\frac{ik}{2\pi \rho^{2}} \int_{\Sigma} (E^{+}-e^{+})\wedge (E^{-}- e^{-})\right)  
Z_{c}(\rho^{-1} E).
\ee
We can use the  the conformal anomaly equation to extract the $\rho$ dependence in $Z_{c}(E)$ hence 
\be\nonumber
\Psi\left(\frac{e^{\pm}}{\rho}\right)=
\exp\left(  \frac{ik}{4\pi \rho^{2}}\int_{\Sigma} e \right)
\rho^{\frac{ic}{24\pi} \int_{\Sigma} e R }
\int D\left(\frac{E}{\rho}\right) \,
\exp\left( 
-\frac{ik}{2\pi \rho^{2}}  \int_{\Sigma} (E^{+}-e^{+})\wedge (E^{-}- e^{-})\right)  
Z_{c}( E)
\ee
In the limit $\rho \to 0$ the term in the integrand become a delta functional imposing $E=e$. It is convenient to make the 
change of variable $E \to e+\rho E$ hence we obtain
\be\nonumber
\Psi\left(\frac{e^{\pm}}{\rho}\right)=
\exp\left(  \frac{ik}{4\pi \rho^{2}}\int_{\Sigma} e \right)
\rho^{\frac{ic}{24\pi} \int_{\Sigma} e R }
\int D{E} \,
\exp\left( 
-\frac{ik}{2\pi}  \int_{\Sigma} E^{+}\wedge E^{-} \right)  
Z_{c}( e +\rho E)
\ee
thus in the limit 
we get 
\be
\Psi\left(\frac{e^{\pm}}{\rho}\right)\sim{\cal N}
\exp\left(  \frac{ik}{4\pi \rho^{2}}\int_{\Sigma} e + \ln\rho \frac{ic}{24\pi} \int_{\Sigma} e R\right)
Z_{c}(e)
\ee
which agree with the holographic renormalisation in the semi-classical limit $c\sim 6k $. 
Here $ {\cal N} = \int D{E} \,
\exp\left( 
-\frac{ik}{2\pi}  \int_{\Sigma} E^{+}\wedge E^{-} \right)=1  $.

Another intriguing limit  to consider, is the opposite limit 
$\lim_{\rho\to 0} \psi(\rho e)$ which amounts to look at the 
value of $\psi$ for the singular metric $e=0$.
In this case it is convenient to introduce the parameterization 
$$e^{+} = e^{\varphi +\alpha}( du + \mu d\tilde{u}),\quad 
e^{-} = e^{\varphi -\alpha}( du + \tilde{\mu} d\tilde{u})
$$
In this limit, the dependence on $e$ formally drops out of the integral 
and the resulting integral is invariant under 2d diffeomorphism 
and lorentz gauge transformation. 
This means that this limit is singular and we can factor out an infinite gauge volume factor  
$V = \mathrm{Vol}(Diff_{\Sigma}) \mathrm{Vol}(Lorentz)$.
The resulting integral contains an integral over the Liouville field and 
the moduli $m$. It is given by 
\be
\Psi(\rho e^{\pm}) \sim_{\rho=0} V \int D\phi Dm D{\bar{m}}\,\,
e^{i \frac{(c-26)}{24 \pi} (S_{L}(\phi,g_m) + \mu \int \sqrt{g_{m}})}
Z_{c}(g_{m})
\ee
with $ (c-26)\mu \equiv 24k$.
The  integral is exactly the definition of  non-critical string theory associated with the CFT $Z_{c}$.
Of course, since the prefactor is infinite this is a extremely singular limit but it is still interesting to see non critical string theory arising from an 
attempt to look at  the wave  function deeply in the bulk.
It would be interesting to understand what is the meaning  of  this property.

\section{Gravity in the first order formalism}\label{1order}
Since in the proof of (\ref{main}) we use the first order formulation of gravity, we review
in this section  the formulation of gravity in the first order formalism and show its equivalence at the hamiltonian level with the WdW equation discussed in the bulk of the paper.
The action of gravity in the metric formalism is given by
\be
S= \frac1{16\pi G} \int_{M} \sqrt{g}\left(R(g) +\frac 2{l^{2}}\right) + 
\frac1{8\pi G}\int_{\partial M} \sqrt{g} K
\ee
In the first order formalism ( Cartan-Weyl formulation of gravity) the dynamical variables are
a  frame field $e^{i}$ and an $SL(2,\mathbb{R})$ connection $\omega^{i}$
and the bulk action is given by  
\be\label{S1}
S_{G}=\frac{1}{8\pi G} \int_{M} e_{i}\wedge R(\omega)^{i} + \frac\Lambda6 \epsilon_{ijk}e^{i}\wedge e^{j}\wedge e^{k}
\ee
where $R^{i}(\omega)=d\omega^{a} + \epsilon^{abc}\omega_{b}\wedge \omega_{c}$, $\Lambda = -1/\ell^{2}$, and the metric is given by 
$ds^{2}= e^{i}\eta_{ij}e^{j}$.
From this Cartan-Weyl formulation, one recover the second order formulation
 if one solves for the connection.
In order to do the hamiltonian analysis, we introduce a `time' 
slice and coordinates $t,u=x+y,\tilde{u}=x-y$.

\be
S_{G}=\frac{1}{8 \pi G} \int {\rm{d}}t \int_{\Sigma} 
\tr \left(e_{u} \partial_{t} \omega_{\tilde{u}} - e_{\tilde{u}} \partial_{t} \omega_{u}
+2 e_{t}\,H(e,\omega)
+\omega_{t}\, G(e,\omega)\right)  \rd u\rd \tilde{u}
\ee
where the constraints are 
\bea
H(e,\omega)&\equiv& \partial_{\tilde{u}}\omega_{{u}}-\partial_{{u}}\omega_{\tilde{u}}- [\omega_{u},\omega_{\tilde{u}}] - \Lambda  [e_{u},e_{\tilde{u}}]\\
G(e,\omega) &\equiv& \partial_{u}e_{\tilde{u}} -\partial_{\tilde{u}}e_{u}
+[\omega_{u},e_{\tilde{u}}] - [\omega_{\tilde{u}}, e_{u}]
\eea
where $e=e^{i}T_{i}$, with $[T_{i},T_{j}] =\epsilon_{ijk} T^{k}$, $\mathrm{tr}(T_{i}T_{j})= \frac12 \eta_{ij}$ (-++) signature.

What we remark first is the fact that there are three lie algebra 
elements $e^{i}$ parameterizing the 2d metric 
$g_{ab} = e^{i}_{a}e^{j}_{b}\eta_{ij}$ on $\Sigma$.
In order to make contact with the hamiltonian formulation 
in terms of metric variable 
we  introduce a Cartan decomposition of the $SL(2,\mathbb{R}) $
algebra, where the generators are $T_{3}=\frac12 \sigma_{3}$, $T_{\pm}=\frac{1}{\sqrt{2}} \sigma_{\pm}$ which satisfies 
$[T_{3}, T_{\pm}] = \pm T_{\pm},\quad [T_{+},T_{-}]= T_{3}$
and 
$\tr(T_{3}T_{3}) = \frac12= \tr(T_{+}T_{-})$.
We decompose the metric in terms of a vector $n\equiv e^{3}$ and a 2d frame field $e^{\pm}$
$$
\frac{e_{\mu}}{\ell}\equiv n_{\mu} T_{3} + e_{\mu}^{+} T_{+} +e^{-}_{\mu}T_{-}.
$$
Similarly we decompose the connection in terms of a $U(1)$ connection $\omega$ and the variables conjugated to $e^{\pm}$.
More precisely we denote 
$$\omega_{\mu} \equiv  \omega^{3}_{\mu},\quad 
\pi^{u}_{\pm}  \equiv \omega_{\tilde{u}}^{\mp},\quad 
\pi^{\tilde{u}}_{\pm}  \equiv -\omega_{u}^{\mp}.
$$
The commutators are given by 
\bea
[\pi^{\mu}_{a}(x), e^{b}_{\nu}(y)] &=& \delta^{b}_{a}\delta^{\mu}_{\nu}\frac{4\pi}{i k} \delta^{(2)}(x-y),\\
{[}\omega_{\tilde{u}}(x), n_{u}(y)] &=& \frac{4\pi}{i k} \delta^{(2)}(x-y),\quad
[\omega_{u}(x), n_{\tilde{u}}(y)] =  - \frac{4\pi}{i k} \delta^{(2)}(x-y)
\eea
where $a,b=\pm$ and $\mu=u,v$.

We can write the constraints in terms of these variables, one obtains
\bea
G^{3} &=&\pi_{+}^{\mu} e_{\mu}^{+} - \pi_{-}^{\mu} e_{\mu}^{-} 
+\partial_{u}n_{\tilde{u}} -\partial_{\tilde{u}}n_{u}\label{G1}\\
G^{+} & =& \nabla_{u}e^{+}_{\tilde{u}} -\nabla_{\tilde{u}}e^{+}_{u} + \pi_{-}^{\mu}n_{\mu}\label{G2}\\
G^{-} & =& \nabla_{u}e^{-}_{\tilde{u}} -\nabla_{\tilde{u}}e^{-}_{u} - \pi_{+}^{\mu}n_{\mu}\label{G3}
\eea
where we have introduce a $U(1)$ covariant derivative 
$$\nabla_{\mu} e_{\nu}^{\pm} \equiv \partial_{\mu}e_{\nu}^{\pm} \pm
\omega_{\mu} e_{\nu}^{\pm}, \quad \nabla_{\mu} \pi^{\nu}_{\pm} \equiv \partial_{\mu}\pi^{\nu}_{\pm} \mp
\omega_{\mu} \pi^{\nu}_{\pm}$$
These represents as we will see generators of $SL(2,\mathbb{R}) $
gauge transformations.
We also have 
\bea 
-H^{\pm} =\nabla_{\mu}\pi^{\mu}_{\mp} - n_{u}e^{\pm}_{\tilde{u}} + n_{\tilde{u}}e^{\pm}_{u}
\eea
which generates two dimensional diffeomorphism and finally 
the hamiltonian constraint
\be
H^{3} =  -(\pi_{+}^{u}\pi_{-}^{\tilde{u}}- \pi_{+}^{\tilde{u}}\pi_{-}^{u})
+  (e_{u}^{+}e_{\tilde{u}}^{-} -e_{\tilde{u}}^{+}e_{u}^{-}) + R(\omega) 
\ee
where $R(\omega) = \partial_{\tilde{u}}\omega_{u}-\partial_{u}\omega_{\tilde{u}} $ is the curvature of the $U(1)$ connection.

This form of the constraints are not exactly the same as the
one in the metric formalism. The reason being that there is an extra pair of canonical variables $n_{\mu},\omega_{\mu}$ compare to 
the metric formulation.  When the covector $n_{\mu}=e^{3}_{\mu}$ is not equal to zero
 the 2-dimensional metric induced on $\Sigma$ is not
  $ds^{2}= e^{+}e^{-}$ but it is  given  by
 \be
 ds^{2}= (e^{+}_{(\mu} e^{-}_{\nu)} + \frac12 n_{\mu}n_{\nu})dx^{\mu}d{x^\nu}
 \ee
However there are also an additional sets of constraints $G^{\pm}$ which generates $SL(2,\mathbb{R})$ gauge transformations of the triplet 
$e^{+},e^{-},n$:
Lets define $G(X) \equiv \int (X_{+} G^{+} +X_{-}G^{-})$, we can compute its action on $e,n$
\bea
[G(X), e^{+}_{\mu}] = -\frac{4\pi}{i k}\, n_{\mu} X_{-},& &
[G(X), e^{-}_{\mu}] = \frac{4\pi}{i k} n_{\mu} X_{+},\\
{[}G(X), n_{\mu}]  &=&  \frac{4\pi}{i k} (e_{\mu}^{-} X_{-}- e_{\mu}^{+} X_{+}  ).
\eea
Using these transformations  and if we now choose the restriction that the metric is time-like we can always go to a Lorentz frame in which $n_{\mu}=0$, which we call the radial gauge.

One may be surprised here that we choose the 2d metric to be timelike since this is unusual but it is perfectly consistent, nothing in the formalism prevent us from choosing at will the signature of the induced metric on $\Sigma$ since this is just a restriction on the 
field configurations which is consistent with the dynamics.
In order to recover the usual hamiltonian case for which $\Sigma$ is spacelike  we just have to chose a basis of generators where $T_{3}$ is timelike.
This is achieved by making the replacements
 $ n_{\mu}\to i n_{\mu}$, $\omega_{\mu}\to i\omega_{\mu} $ together with the replacement of the null coordinates $u,\tilde{u}$ by complex coordinates $u=z, \tilde{u}=\bz$ and the change of canonical momenta
$\pi^{\pm}_{\mu} \to i \pi^{\pm}_{\mu}$.
In this case $e^{+}_{z}$ and $e^{-}_{\bz}$ are complex conjugate.
Note that we could also consider the case of Riemannian gravity 
which just correspond to a Wick rotation of the coordinates $u,\tilde{u} \to z, \bz$.
In the following we nevertheless continue to deal with the Lorentzian AdS case where $u,\tilde{u}$ are real null coordinates and $ e^{\pm}_{u},e^{\pm}_{\tilde{u}}$ are real fields.

In order to make the link with the metric formulation we therefore have to chose the time gauge $n_{\mu} =0$.
That is suppose that we have a wave functional $\Psi(e^{+},e^{-},n)$ 
which satisfy the constraints $G^{+}\Psi =G^{-}\Psi =0$ we can then defined a functional $\tilde{\Psi}(e^{+},e^{-}) \equiv \Psi(e^{+},e^{-},n)|_{n=0}$.
In order to implement this gauge we have however to 
compute what is the residual action  of $\omega$, which is  the field conjugated to $n$, on $\tilde{\Psi}$.
In order to do so we define the spin connection $\omega_{\mu}(e)$ 
which satisfies $de^{\pm} \pm \omega\wedge e^{\pm}=0$ or explicitly
\bea
\partial_{u}e^{+}_{\tilde{u}} -\partial_{\tilde{u}}e^{+}_{u} 
&=& \omega_{\tilde{u}}(e)e^{+}_{u} -\omega_{u}(e)e^{+}_{\tilde{u}},  \\
\partial_{u}e^{-}_{\tilde{u}} -\partial_{\tilde{u}}e^{-}_{u} 
&=& \omega_{u}(e)e^{-}_{\tilde{u}} -\omega_{\tilde{u}}(e)e^{-}_{u}.
\eea
We assume that  the 2d induced metric is invertible and we denote $e^{\mu}_{a}$ the 2d inverse frame field $ e_{\mu}^{a} e^{\mu}_{b}=\delta^{a}_{b}$.
We can  compute 
\be
e^{\mu}_{+} G^{+} - e^{\mu}_{-} G^{-} 
= \epsilon^{\mu\nu}(\omega_{\nu}- \omega_{\nu}(e)) - n_{\nu}(e_{+}^{\mu} \pi_{-}^{\nu}+e_{-}^{\mu} \pi_{+}^{\nu})
\ee
Taking the action of this operator on a gauge invariant functional satisfying the constraints  $G^{+}\Psi =G^{-}\Psi =0$ one conclude 
\be
\left. \omega_{\mu}\Psi (e^{+},e^{-} , n)\right|_{n=0}
=\left. \frac{4\pi}{i k} \epsilon_{\mu\nu} \frac{\delta}{\delta n_{\nu}}\Psi (e^{+},e^{-} , n)\right|_{n=0}
=\omega_{\mu}(e)\Psi (e^{+},e^{-} , 0)
\ee
Not surprisingly one obtains that in the radial gauge the action of $\omega_{\mu}$ is therefore given by multiplication 
by the spin connection.

One has a residual gauge constraints $G^{3}$ which reads
in the radial gauge
\be
(\pi_{+-}-\pi_{-+})\tilde{\Psi}(e^{+},e^{-}) =0
\ee
hence this imply that $\pi$ is a symmetric tensor.
This equation is satisfied if $\Psi$ depends on $e^{\pm}$ only via the 
metric $\bga_{\mu\nu} = e^{+}_{(\mu}e^{-}_{\nu)}$.

Thus in the time gauge and once we solve all the gauge constraints
$G^{a}=0$ the wave equation is a functional of the metric solution of $H^{a}\Psi =0$ which can easily be seen to be just the usual metric diffeomorphism and hamiltonian constraints.

\section{Proof}\label{proof}
We can now present the proof of the reconstruction formula 
(\ref{main}).
Lets consider  the hamiltonian constraint in the first order formalism
\be
\hat{H}  = \hat{T} +  R(e), \quad  \hat{T}\equiv  -(\hat{\pi}_{+}^{u}\hat{\pi}_{-}^{\tilde{u}}- \hat{\pi}_{+}^{\tilde{u}}\hat{\pi}_{-}^{u})
+ (e_{u}^{+}e_{\tilde{u}}^{-} -e_{\tilde{u}}^{+}e_{u}^{-})
\ee
where $$\hat{\pi}^{u}_{\pm} = \frac{4\pi}{i k} \frac{\delta}{\delta e^{\pm}_{u}},\quad \quad\hat{\pi}^{\tilde{u}}_{\pm} = \frac{4\pi}{i k} \frac{\delta}{\delta e^{\pm}_{\tilde{u}}} $$ and $R(e) = R(\omega(e))$.
One note here that there is no operator ordering ambiguity in the definition of $\hat{T}$.
One introduce a notation for the kernel entering the integral formula (\ref{main})
\be
K(e,E)\equiv \exp\left( 
-\frac{ik}{2\pi} \int_{\Sigma} (E^{+}-e^{+})\wedge (E^{-}- e^{-})\right)
\ee

On decomposes the proof if two parts:
One looks separately at how the  action of  $\hat{T}$ and the action of $R(e)$  on $\Psi(e)$ is reflected onto $Z_{c}(E)$. 

\subsection{Kinetic term}
One starts with the action 
$$
\hat{\pi}^{u}_{\pm} \int \det(e) = \pm \frac{4\pi}{ik }  \,  e_{\tilde{u}}^{\mp},
\quad 
\hat{\pi}^{u}_{\pm} \int \det(e) =  \mp \frac{4\pi}{ik }  \,e_{{u}}^{\mp},
$$
where $\det(e) = e^{+}_{u}e^{-}_{\tu} -e^{+}_{\tu}e^{-}_{u}$.

One can commute the action of $\hat{\pi}_{+}^{u}\hat{\pi}_{-}^{\tilde{u}}$ through the first term in (\ref{main})
\beq\nonumber
 \left[\hat{\pi}_{+}^{u}\hat{\pi}_{-}^{\tilde{u}} - e_{u}^{+}e_{\tilde{u}}^{-}\right] \Psi\left(e\right)& =&
\exp\left( \frac{ik}{4\pi}\int_{\Sigma} \det(e) \right)
\int DE \, Z_{c}(E)
\\
& &\times \left[\left( \hat{\pi}_{+}^{u} +  e_{\tilde{u}}^{-}\right)
\left( \hat{\pi}_{-}^{\tilde{u}}+  e_{{u}}^{+}\right)- e_{u}^{+}e_{\tilde{u}}^{-}\right]K(e,E) \eeq
we expand the operator as follows
\be\label{pipi}
  \left[\left( \hat{\pi}_{+}^{u} +  e_{\tilde{u}}^{-}\right)
\left( \hat{\pi}_{-}^{\tilde{u}}+  e_{{u}}^{+}\right)- e_{u}^{+}e_{\tilde{u}}^{-}\right]
=
\frac12 \hat{\pi}_{+}^{u}\left( \hat{\pi}_{-}^{\tilde{u}}+ 2 e_{{u}}^{+}\right)
+ \frac12 \hat{\pi}_{-}^{\tilde{u}}\left( \hat{\pi}_{+}^{u} + 2 e_{\tilde{u}}^{-}\right) -[\hat{\pi}_{-}^{\tilde{u}}, e_{\tilde{u}}^{-} ]
\ee
We can now use the following identities to convert differential operator 
of $e$ into differential operators acting on $E$
$$ 
 \left( \hat{\pi}_{-}^{\tilde{u}} + 2 e_{{u}}^{+} \right)K(e,E)= 2 E^{+}_{{u}} K(e,E)
$$
$$
 \left( \hat{\pi}_{+}^{{u}} +\frac{4\pi}{ik}\frac{\delta}{\delta E_{u}^{+}}\right)K(e,E)= 0
$$
Doing the same manipulation for the second term in (\ref{pipi})
one gets a contribution 
$ -\frac{4\pi}{ik}\left[E_{u}^{+} \frac{\delta}{\delta E_{u}^{+}} 
+ E_{\tilde{u}}^{-} \frac{\delta}{\delta E_{\tilde{u}}^{-}} 
\right] K(e,E).
$
If one symmetrize this expression one need to commute 
some terms   $ {\delta}/{\delta E_{u}^{+}}$ with $E_{u}^{+}$,
this commutators cancels {\it exactly} the third   term in (\ref{pipi}).
We are then left with
\beq\nonumber
 & &\left[\hat{\pi}_{+}^{u}\hat{\pi}_{-}^{\tilde{u}} - e_{u}^{+}e_{\tilde{u}}^{-}\right] \Psi\left(e\right)=
\exp\left( \frac{ik}{4\pi}\int_{\Sigma} \det(e) \right)
\int DE \,Z_{c}(E)
 \times \\
& & -\frac{2\pi}{ik}\left[E_{u}^{+} \frac{\delta}{\delta E_{u}^{+}} 
+ E_{\tilde{u}}^{-} \frac{\delta}{\delta E_{\tilde{u}}^{-}} 
+ \frac{\delta}{\delta E_{u}^{+}} E_{u}^{+}
+ \frac{\delta}{\delta E_{\tilde{u}}^{-}}  E_{\tilde{u}}^{-}
\right]
 K(e,E)
\eeq
Doing similar manipulations for the operator 
$\hat{\pi}_{+}^{\tilde{u}}\hat{\pi}_{-}^{u} -e_{\tilde{u}}^{+}e_{u}^{-}$
leads to the conclusion 
\beq\nonumber
 \hat{T} \Psi\left(e\right)=
\exp\left( \frac{ik}{4\pi}\int_{\Sigma} \det(e) \right)
\int DE \,Z_{c}(E)\,
  \frac{2\pi}{ik}\left[\frac{\delta }{\delta \phi} - \left(\frac{\delta}{\delta \phi} \right)^{\dagger}\right]
K(e,E)
\eeq
where we denote the infinitesimal conformal rescaling 
$$\frac{\delta }{\delta \phi} \equiv \left[E_{\mu}^{a} \frac{\delta}{\delta E_{\mu}^{a}}\right],
\quad\quad
-\left(\frac{\delta}{\delta \phi} \right)^{\dagger} = \left[\frac{\delta}{\delta E_{\mu}^{a}} E_{\mu}^{a} \right]
$$

The measure of integration is not invariant under rescaling, following the argument of David-Distler-Kawai \cite{DavidDistler}, and accordingly $\frac{\delta}{\delta \phi}$ is 
not anti-hermitic.
Lets  introduce the parameterization 
$$
e^{+} = e^{\varphi +\alpha}\hat{e}^{+}_{\mu},\quad \quad \hat{e}^{+}_{\mu}\equiv( du + \mu d\tilde{u}),\quad 
e^{-} = e^{\varphi -\alpha}\hat{e}^{-}_{\tilde{\mu}}, \quad\quad \hat{e}^{-}_{\tilde{\mu}} \equiv ( du + \tilde{\mu} d\tilde{u})
$$
The measure can be split into 
$DE = D\phi D\alpha D\hat{g}_{\mu, \tilde{\mu}}$.
the problem is that the measure for $\phi$ and $\alpha$ depends 
non linearly on $\phi$
\be
(\delta e,\delta e) = \int \hat{e} e^{2\phi} (\delta\phi^{2} - \delta{\alpha}^{2})\ee
if one focus on variation of $\phi,\alpha$ only.
The idea of David, Distler and Kawai \cite{DavidDistler} 
is to express this measure in terms of 
a translation invariant measure in $\phi$ and $\alpha$ that we denote $D_{0}\phi$.
What as been show by computing the jacobian of the transformation and invoking self consistent condition \cite{Liouville} is that we can express the original measure in terms of a translationally invariant measure  provided we introduce a jacobian which is proportional to the exponential of the Liouville action. That is 
\be
DE = D_{0}\phi D_{0}\alpha Dg_{\mu, \tilde{\mu}}\, \exp\left(-{\frac{2 i}{24\pi} S_{L}(\phi, g_{\mu,\tilde{\mu}})}\right)
\ee
The factor $2$ comes from the fact that both $\phi$ and $\alpha$ contributes to the jacobian.
This imply that  $\delta /\delta \phi$ is not an hermitian operator but
such that 
\be
\left(\frac{\delta}{\delta \phi} \right)^{\dagger} = -\frac{\delta}{\delta \phi} + \frac{i}{12\pi} R(E)
\ee
We finally obtain
\beq\nonumber
 & &\hat{T} \Psi\left(e\right)=
\exp\left( \frac{ik}{4\pi}\int_{\Sigma} \det(e) \right)
\int DE \, \,
K(e,E)\, 
\frac{4\pi}{ik} \left(-\frac{\delta }{\delta \phi} + \frac{i}{24\pi} R(E) \right)Z_{c}(E)
\eeq

\subsection{the potential}

We now turn to the action of the potential term on the wave function $\Psi$.
The curvature is $$R(e) = 
\partial_{\tilde{u}}\omega_{{u}}(e)
 -\partial_{u}\omega_{\tilde{u}}(e)$$
where the spin connection is 
\beq
\det(e)^{-1}\left(e^{-}_{\tilde{u}} \rd e^{+} + e^{+}_{\tilde{u}}\rd e^{-}\right)= \omega_{\tilde{u}}(e) \\
\det(e)^{-1}\left(e^{-}_{{u}} \rd e^{+} + e^{+}_{{u}}\rd e^{-}\right) = \omega_{{u}}(e) \\
\ \partial_{u}e^{\pm}_{\tilde{u}}- 
\partial_{\tilde{u}}e_{u}^{\pm} \equiv \rd e^{\pm} 
\eeq
Let us define the operator 
$$\hat{\Pi} _{\mu}^{a}
\equiv \frac{2\pi}{ik} 
\epsilon^{ab}\epsilon_{\mu \nu}
\frac{\delta }{\delta E_{\nu}^{b}}$$ with 
$\epsilon^{ab}, \epsilon_{\mu \nu}$ antisymmetric tensors 
normalized by $\epsilon^{+-} =1$, $\epsilon_{u\tilde{u}}=1$.
One can check that the operators $ E_{\mu}^{a} + \hat{\Pi}_{\mu}^{a} $ all commute with each others.
Moreover, their action on the kernel gives
\be\nonumber
\left( E_{\mu}^{a} + \hat{\Pi}_{\mu}^{a} 
\right)K(e,E)= e_{\mu}^{a}
K(e,E)
\ee
therefore the action of $R(e)$ on $\Psi(e)$ is equivalent to the action of  $R(E+\hat{\Pi})$ on the kernel.
From the definition of the spin connection and being careful with the operator ordering one can express the difference 
\beq\label{omega}
\omega_{\mu}(E+\hat{\Pi}) -\omega_{\mu}(E) &=& 
\left( (\nabla \hat{\Pi}^{+}) E^{-}_{\mu} + (\nabla \hat{\Pi}^{-}) E^{+}_{\mu}\right)\det(E+\hat{\Pi})^{-1},\\
\quad \nabla \hat{\Pi}^{\pm} &\equiv& \rd \hat{\Pi} \pm \omega(E)\wedge \hat{\Pi}^{\pm}.
\eeq
One might worried that the inverse of $\det(E+\hat{\Pi})$ is a dangerous operator however one has to remember that 
 it is acting on the exponential kernel and its action is just equal to 
 the inverse of $\det(e)$ which is well defined.
 To go further we introduce $ \hat{D}^{\pm}_{\mu} \equiv\mp E^{\pm}_{\mu} \nabla \Pi^{\mp}$ which are the generators of chiral diffeomorphisms:
  \be
\int \xi^{\mu}\hat{D}^{\pm}_{\mu} E^{\pm}_{\alpha} =
{\cal L}_{\xi} E^{\pm}_{\alpha} \pm \xi^{\mu}\omega_{\mu}(E)E^{\pm}_{\alpha}
\ee
where ${\cal L}_{\xi}$ is the Lie derivative and the second term represent an infinitesimal  Lorentz gauge transformation
with parameter $\xi^{\mu}\omega_{\mu}$.
In the RHS of (\ref{omega}) we recognise the generators of diffeomorphism up to a commutator
 $$
 [\nabla \Pi^{+}(x), E_{\mu}^{-}(y)] = - \frac{2\pi}{ik}\nabla_{\mu}^{+} \delta^{(2)}(x,y),\quad 
  [\nabla \Pi^{-}(x), E_{\mu}^{+}(y)] =   \frac{2\pi}{ik} \nabla_{\mu}^{-} \delta^{(2)}(x,y)
 $$
 with $\nabla^{\pm}_{\mu} =\partial_{\mu} \pm \omega_{\mu}$.
 
 Thus overall and after integration by part we have 
 \beq
 \omega_{\mu}(e) \Psi(e)
 =
\exp\left( \frac{ik}{4\pi}\int_{\Sigma} \det(e) \right)
\left(
\int DE \, \,
K(e,E)\,\omega_{\mu}(E) Z_{c}(E)  \right.\\
+\left. \int DE \, \,
 K(e,E) \det(e)^{-1} \left(\hat{D}_{\mu}^{+} -\hat{D}_{\mu}^{-} + R_{\mu}\right)Z_{c}(E) \right) \\
\eeq
where $ \hat{D}^{\pm}_{\mu} \equiv E^{\pm}_{\mu} \nabla \Pi^{\mp}$ are the generators of chiral diffeomorphism 
Since $Z_{c}(E)$ is invariant under Lorentz gauge transformation and 
diffeomorphisms it is annihilated by the action of $\hat{D}^{\pm}_{\mu}$.

The last term is 
a commutator term
\beq
R_{\mu}&=& \left(  E^{-}_{\mu}\left( (\nabla \hat{\Pi}^{+})^{\dagger} - (\nabla \hat{\Pi}^{+})\right)+  E^{+}_{\mu}\left( (\nabla \hat{\Pi}^{-})^{\dagger} - (\nabla \hat{\Pi}^{-})\right)\right)
\eeq
The operator $\int X_{a}\nabla \hat{\Pi}^{a}$ generates 
transformations 
\be
\int X_{a}\nabla \hat{\Pi}^{a} E^{\pm}_{\mu} \equiv \delta_{X} E^{\pm}_{\mu} 
= \nabla_{\mu}X^{a}
\ee
the key point is that the distance (\ref{dist}) we have chosen to 
define the integration measure is invariant under such transformations
\beq
\delta_{X}(e,e) &=&  \int \delta_{X}e^{+}\wedge e^{-} + 
 \int e^{+}\wedge \delta_{X}e^{-}\\
 &=&   -\int {X}^{+} \nabla\wedge e^{-} 
-\int {X}^{-} \nabla\wedge e^{+} =0
\eeq
This means that the measure is invariant under such transformation hence $\nabla \hat{\Pi}^{a}$ is an hermitian operator, thus 
$R_{\mu}=0$,which is the result we wanted to establish.
This shows that 
\beq\nonumber
 & &\hat{H} \Psi\left(e\right)=
\exp\left( \frac{ik}{4\pi}\int_{\Sigma} \det(e) \right)
\int DE \, \,
K(e,E)\, 
\frac{4\pi}{ik}\left(- \frac{\delta }{\delta \phi} + \frac{i}{24\pi}\left(6k +1\right) R(E) \right)Z_{c}(E)
\eeq

%

%

\section{Conclusion}

In this work we have given a new perspective on the relationship between quantum gravity in the bulk and CFT on the boundary
in the context of a non zero cosmological constant.
As we have seen the AdS-matrix of gravity satisfies a radial 
Wheeler de Witt equation which is  a second order equation.
One has derived that the asymptotic of any solution 
of the radial WdW equation is given by a pair of CFTs.
We have argued that restricting to one CFT  amount to look at 
quantum spacetime which possess an asymptotic infinity.
We have shown that the CFT can be interpreted as an initial value data for a radial quantum gravity state. Thus the AdS/CFT correspondence is one to many, that is it is a correspondence between radial states of quantum gravity  and different CFTs.
An explicit way to see this, is to give an explicit reconstruction formula
of the radial quantum gravity state from the boundary CFT. This has been achieved in the context of 2+1 quantum gravity.
From this work we can to some extent answer all the questions asked in the introduction.

Our work suggests several avenues 
which would need to be resolved in order to have a deeper understanding of the AdS/CFT correspondence.
 For instance, as we discussed in section \ref{grav}, it is far from obvious
  that there is a clear  relationship between Euclidean and Lorentzian bulk gravity solutions; 
  except when these solutions are evaluated at on a static boundary metric;
   since the usual correspondence between Euclidean and Lorentzian solution of Wheeler-deWitt equation do not 
 hold anymore for the radial solution of WdW.  
 However since one can always defined a Wick rotated boundary CFT, assuming it is unitary,
 one expect a correspondence between these solutions via the Boundary CFT.
 It would be extremely interesting to find out what is explicitely this mapping at the quantum gravity level.

 Also in our derivation of the asymptotic form of
 the radial WdW equation we have argued that 
  only a finite number of terms should be renormalised in the asymptotic expansion and we have shown this explicitely in the case of 2+1 dimensions. 
  It would be however quite instructing to see how this work exactly by carrying out the Schr\"odinger renormalisation program of Symanzik 
 in the context of gravity.
 
 One open problem that arises here, concerns the meaning of the radial state of gravity. More precisely, it would be extremely useful to unravel the relationship between radial state and usual states of quantum gravity. One expect such a radial state to correspond to a gravitational density matrix, can we find the explicit form of this correspondence?
 Does a radial state, hence the CFT,  allows us to reconstruct the Hilbert space of quantum gravity or does it knows only about 
 some specific subset of states?

 We have given a prescription for the construction of {\it the} CFT associated with quantum gravity in the Lorentzian context. This needs to be further developed and one should do it explicitly in the context of 3d gravity.
 Also as we have seen, the asymptotic expansion shows that one expects to have a CFT corresponding to such a quantum gravity radial state.  However we do not know wether this CFT is irreducible 
 or given by a sum of irreducible CFT. This should be resolved.
 
 Most of our analysis have been done in the context of AdS, but as we emphasized several time most of our results are also valid in the dS context.
 In that case we obtain a CFT partition function associated with an imaginary central charge. That is even if the boundary metric is Riemannian the CFT correspond to a quantum weight $e^{iS}$ and not a statistical weight $e^{-S}$.
 How can we interpret such a ``imaginary'' CFT, How is this related to 
 the thermal properties of de Sitter space? These questions certainly needs to be developed further.
 
 It would be interesting to understand the appearance of the non critical string amplitude in the singular limit of the reconstruction
 formula given in 2+1.
 Finally It would be extremely interesting to propose a bulk reconstruction formula analogous to (\ref{main}) in the context of 4 dimensions.
 Of course a full quantum gravity one is hopeless at first sight but a semi-classical one might be within reach.
We hope to come back to this issue in the near future. 

\vspace{6 pt}
{\bf Acknowledgments:} First, I  would like to thank 
R. Myers  who strongly motivated me  to work on this problem.
I also benefited greatly from discussions with A. Maloney, L. Susskind, J. Gomis, J. Maldacena, L. Smolin, K. Krasnov, R. Leigh, D. Minic and the participants of the 3d quantum gravity workshop in Montreal and the loops and Foam meeting in Zakopane where this work was presented.
\section{Appendix}

\subsection{One loop renormalisation}
In order to perform the renormalisation of the hamiltonian at one loop 
we need to compute the divergent part of the boundary to boundary propagator at one loop.
In order to do so lets start from the formal path integral.
\be\label{S-matrix1}
\Psi_{{\Sigma}}(\bga) =\int_{g|_{\partial M}= \bga}\!\!\!\!\!\!\!\mathrm{D} g \,\,e^{i S_{M}(g)}  
\ee
In order to define this path integral 
we first split the bulk metric in terms of a background metric and a perturbation $g= \bar{g} +h $. $\bar{g}$ is a solution of the classical equations of motion 
$$R_{\alpha \beta}(\bar{g})= -\frac{d}{l^{2}} \bar{g}_{\alpha\beta},\quad \bar{g}_{ab}|_{\partial M}= \gamma_{ab},\quad  \bar{g}_{0b}= \delta_{0b}$$
where $0$ refer to the coordinate transverse to the boundary.
$h$ is the quantum field that we need to integrate over. It is such that its components parallel to the boundary satisfies Dirichlet boundary conditions $h_{ab}|_{\partial M}=0$. The components transverse to the boundary are not restricted and should be integrated over.
This transverse integration implement the constraints.
In order to make sense of the path integral we also need to perform a gauge fixing.
The original gauge invariance can be divided between between the background and perturbation either  as  a background gauge invariance
\be\label{gaugeb}
\delta^{B}_{\xi}\bar{g}_{\mu\nu} = 2 \nabla_{(\mu}\xi_{\nu)}, \quad 
\delta^{B}_{\xi} h_{\mu\nu} = {\cal L}_{\xi} h_{\mu\nu}=
\xi^{\rho}\nabla_{\rho} h_{\mu \nu} + 2 h_{\rho (\mu}\nabla_{\nu)}\xi^{\rho}
\ee
or as a gauge invariance
\be
\delta_{\xi}\bar{g}_{\mu\nu} = 0, \quad 
\delta_{\xi} h_{\mu\nu} =   2 \nabla_{(\mu}\xi_{\nu)} + {\cal L}_{\xi} h_{\mu\nu}, \quad \xi|_{\partial M}= 0 \ee
where the covariant derivative is with respect to $\bar{g}$ and all indices are raised with $\bar{g}$.
This gauge invariance of the action needs to be fixed. 
And we denote $F^{\mu}(g)$ the corresponding gauge condition 
and $J^{\mu}_{\nu}$ its variation with respect to diffeomorphisms
$$\delta_{\xi} F^{\mu} = J^{\mu}_{\nu} \xi^{\nu}.$$
In order to respect the background diffeomorphism symmetry 
such a gauge should be invariant under the background gauge transformation $\delta^{B}_{\xi} F^{\mu}\sim F^{\nu}$.
A convenient gauge is for instance the deDonder gauge 
\be
F^{\mu}(g;h)= \nabla_{\nu}h^{\mu \nu} - \frac12 \nabla^{\mu} h,\quad 
J^{\mu}_{\nu} = \Box\delta^{\mu}_{\nu} + R^{\mu}_{\nu} +O(h)
\ee
The terms of order $h$ will not contribute to one loop.
Such a gauge fixing is taken into account by inserting the Faddev-Popov term 
\be
\prod_{x,\mu}\delta(F^{\mu}(x)) \det(J^{\mu}_{\nu}(x)) \,\,\,\,\,\mathrm{or}\,\,\,\,\,
\frac{\det(J)}{(\det{i \alpha \bar{g}})^{\frac12}} 
e^{i  \frac\alpha2 \int_{M}\sqrt{\bar{g}} F^{\mu}\bar{g}_{\mu\nu}F ^{\nu}}
\ee
or more conveniently by taking an average 
\be
\int D\alpha^{\mu} e^{-\frac12\int\sqrt{\bar{g}} \alpha^{\mu}\bar{g}_{\mu\nu}\alpha^{\nu}} \prod_{x,\mu}\delta(F^{\mu}(x) -\alpha^{\mu}(x) ) \det(J)
\ee
We can now, following Barvinsky \cite{Barv1}, define the 
One loop corrected gravity wave function.
We denote the bulk operator denoted $\Box^{\alpha\beta; \mu\nu}= \delta^{2} S/\delta g_{\alpha \beta}\delta g_{\mu\nu}$.
The Green function $G_{(ND)}$ entering the definition 
of the one loop partition function satisfies mixture of Dirichlet with Neuman Boundary conditions. 
the component tangential to the boundary satisfy Diriclet boundary conditions while the comonent transverse to the boundary satisfy Neumann conditions induced by the 
gauge fixing.
Namely, denoting $a,b$ 
indices tangential to the boundary we have
\beq
\Box^{\alpha\beta; \mu\nu} G_{\mu\nu; \rho \sigma}(x,y) 
&=&\delta^{\alpha \beta}_{\rho \sigma}\delta(x,y) \\
G_{ab; \mu\nu}(x,y)|_{x\in\partial M} &=&0 \\
\overrightarrow{F}_{\mu}^{\alpha \beta} G_{\alpha \beta; \mu\nu}(x,y)|_{x\in\partial M} &=&0
\eeq  
where the gauge condition is $F_{\mu}(h) \equiv \overrightarrow{F}_{\mu}^{\alpha \beta} h_{\alpha \beta}$.
The one-loop wave functional is given by
\be
\psi_{\Sigma}(\bga) = \frac{\mathrm{Det}_{D}(J)}{\left(\mathrm{Det}_{DN}(\Box)\right)^{\frac12}}
e^{\frac{i}{\kappa}S^{(HJ)}(\bga)}
\ee
where $S^{(HJ)}(\bga)$ is the hamilton-jacobi functional
and $\mathrm{Det}_{DN}(\Box)\equiv \mathrm{Det}^{-1}(G^{(DN)})$ while the ghost determinant is computed with 
Dirichlet conditions.
These determinants should be regularised with the heat kernel methods and are such that  the local singulrr terms are substracted.
In the case of usual scalar field theory one can check that this renormalised wave function do satisfy the 
Schr\"odinger equation modified by the  one loop substraction (\ref{substraction}) of the Schr\"odinger operator \cite{Osborn}.
One expect the same result in the context of the WdW equation but the explicit check as not been performed yet.

\end{document}